\documentclass[iop]{emulateapj}

\usepackage{graphicx}
\usepackage{amsmath}
\usepackage{subfigure}
\usepackage{float}
\usepackage{natbib} 
\usepackage{hyperref}
\usepackage{tabularx}

\begin{document}

\title{EvryFlare I: long-term Evryscope monitoring of flares from the cool stars across half the Southern sky}

\author{Ward S. Howard\altaffilmark{1}, Hank Corbett\altaffilmark{1}, Nicholas M. Law\altaffilmark{1}, Jeffrey K. Ratzloff\altaffilmark{1}, Amy Glazier\altaffilmark{1}, Octavi Fors\altaffilmark{1,2}, Daniel del Ser\altaffilmark{1,2}, Joshua Haislip\altaffilmark{1}}
\altaffiltext{1}{Department of Physics and Astronomy, University of North Carolina at Chapel Hill, Chapel Hill, NC 27599-3255, USA}
\altaffiltext{2}{Dept. de F\'{\i}sica Qu\`antica i Astrof\'{\i}sica, Institut de Ci\`encies del Cosmos (ICCUB), Universitat de Barcelona, IEEC-UB, Mart\'{\i} i Franqu\`es 1, E08028 Barcelona, Spain \newline}
\email[$\star$~E-mail:~]{wshoward@unc.edu}

\begin{abstract}
We search for superflares from 4,068 cool stars in 2+ years of Evryscope photometry, focusing on those with high-cadence data from both Evryscope and TESS. The Evryscope array of small telescopes observed 575 flares from 284 stars, with a median energy of 10$^{34.0}$ erg. Since 2016, Evryscope has enabled the detection of rare events from all stars observed by TESS through multi-year, high-cadence continuous observing. We report $\sim$2$\times$ the previous largest number of 10$^{34}$ erg high-cadence flares from nearby cool stars. We find 8 flares with amplitudes of 3+ \textit{g}\textsuperscript{$\prime$}~magnitudes, with the largest reaching 5.6 magnitudes and releasing 10$^{36.2}$ erg. We observe a 10$^{34}$ erg superflare from TOI-455 (LTT 1445), a mid-M with a rocky planet candidate \footnote{During proofs, a pre-print of a paper by Winters et al. confirming LTT 1445Ab was released. See arxiv.org/abs/1906.10147}. We measure the superflare rate per flare-star and quantify the average flaring of active stars as a function of spectral type, including superflare rates, FFDs, and typical flare amplitudes in \textit{g}\textsuperscript{$\prime$}. We confirm superflare morphology is broadly consistent with magnetic re-connection. We estimate starspot coverage necessary to produce superflares, and hypothesize maximum-allowed superflare energies and waiting-times between flares corresponding to 100\% coverage of the stellar hemisphere. We observe decreased flaring at high galactic latitudes. We explore the effects of superflares on ozone loss to planetary atmospheres: we observe 1 superflare with sufficient energy to photo-dissociate all ozone in an Earth-like atmosphere in one event. We find 17 stars that may deplete an Earth-like atmosphere via repeated flaring. Of the 1822 stars around which TESS may discover temperate rocky planets, we observe 14.6$\pm$2\% emit large flares.
\end{abstract}


\maketitle

\section{Introduction}
Stellar flares occur on main-sequence stars when convection of the photosphere distorts the star's magnetic field, leading to a magnetic re-connection event that releases large quantities of stored magnetic energy. Electrons are accelerated down magnetic field lines toward the photosphere, colliding with and heating plasma to temperatures above 20 MK. The depths at which these electrons brake during the flare determines the wavelengths at which the plasma radiates. White-light flares are thought to result from electron collisions in the photosphere \citep{allred2015}, although additional mechanisms have been proposed (e.g. \citet{Fletcher_Hudson2008,Heinzel_Shibata2018,Jejcic2018}). White-light flares may last from minutes to hours, following a fast-rise and exponential-decay (FRED) profile in time-domain observations, e.g. \citet{davenport2014}. Because flaring depends on the magnetic field of the star, increased flare activity is correlated with young stellar age \citep{Ambartsumian1975,Davenport2019,Ilin2019}, fast stellar rotation (e.g. \citet{west2015,Astudillo-Defru2017,Newton2017,Wright2018}), high starspot coverage \citep{Yang2017}, and late spectral type \citep{West2008,west2015,Wright2018,Davenport2019,Paudel2019}.

Because Earth-sized planets orbiting cool stars are both common \citep{Dressing2013, DressingCharbonneau2015} and produce high-SNR transit and radial velocity signals, cool stars are popular targets in the search for nearby Earth-like exoplanets. Enhanced flaring is often observed from cool stars (i.e. late K-dwarf and M-dwarf stars with T$_\mathrm{eff}<$ 4000 K) \citep{Muirhead2018}; a deep convection zone and fast stellar rotation increase the available magnetic energy and may result in high flare rates and flare energies up to $1000\times$ greater than those observed from the Sun (e.g. \citet{allred2015, davenport2016} and references therein). Not only are cool stars the most populous stellar types in the galaxy \citep{chambrier2003,henry2004,henry2006}, but detecting flares from cool stars is easier than detecting flares from Solar-type stars due to their lower luminosity and higher flare contrast \citep{allred2015, Schmidt_2018_ASASSN_flares}.

Temperate rocky planets have already been discovered in orbit around several nearby cool stars, e.g. \citet{anglada2016,Gillon2017,Dittmann2017,Bonfils2018}. However, intense flaring may pose problems for the habitability of planets orbiting cool stars. The so-called ``habitable zone" (HZ) is defined as the distance from a star at which the stellar flux and planet atmosphere would allow for the existence of liquid water on the surface \citep{Kopparapu2013}. The low luminosity of cool stars requires HZ orbital distances to be very close to the star, resulting in increased flare energy and high-energy stellar particles reaching the planetary atmosphere. Combined with the intrinsically-high flare rate of active cool stars, the ozone layers of Earth-like planets may be suppressed or destroyed on geologically-short timescales \citep{Tilley2019, Howard2018, Loyd2018}. The increased activity of the young Sun altered the atmosphere of the early Earth (e.g. \citet{Cockell2002,Rugheimer2015,Rugheimer2018}) but did not prevent life on our planet. \citet{Airapetian2016} find that superflares may have even increased the habitability of our planet by fixing inert atmospheric nitrogen. Furthermore, the atmospheres of nearby M-dwarf planets may be capable of shielding life from the most extreme UV radiation \citep{OMalley2019arXiv}.

\begin{figure*}
	\centering
	{
		\includegraphics[trim= 1 20 1 15,clip, width=6.9in]{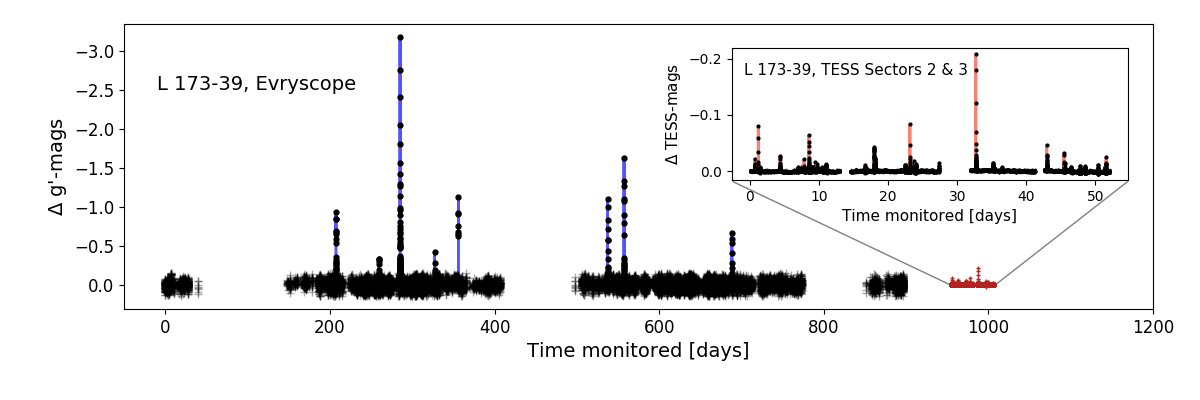}
	}
	\caption{The full 2016-18 Evryscope light curve of flare star L 173-39. This flare star demonstrates how the Evryscope light curve complements the TESS light curve. While 2 sectors of TESS observations captures the frequent flares of lower to moderate energy, long-term Evryscope observations at more moderate photometric precision capture the rare, high-energy flares. Combined, these surveys sample a broader flare distribution from each star.}
	\label{fig:full_lc}
\end{figure*}

High-energy particles that may be associated with large flares deplete atmospheric ozone through the creation of nitrogen-oxide species. While the ozone layer of an Earth-like planet may withstand single superflare events of 10$^{34}$ erg \citep{Segura2010,Greissmeier2016,Tabataba2016}, the cumulative effect of multiple superflare events per year does not allow the planetary atmosphere to recover \citep{Tilley2019}. The largest flares may fully photo-dissociate an ozone column in a single event without consideration of high energy particles at all \citep{Loyd2018}. 

Long-term X-ray and UV flare emission may contribute to the complete stripping away of Earth-like atmospheres \citep{cuntz2016, owen2016}. \citet{luger2015} notes that photoevaporation of mini-Neptune atmospheres may lead to habitable worlds rather than prevent them. However, this outcome is only likely for specific H/He-envelope mass fractions, core sizes, and incident stellar fluxes \citep{owen2016}.

Since July 2018, the Transiting Exoplanet Survey Satellite (TESS, \citealt{Ricker_TESS}) has been searching for transiting exoplanets across the entire sky, split into 26 sectors. Each TESS sector is continuously observed in the red by four 10.5 cm optical telescopes for 28 days at 21$\arcsec$ pixel$^{-1}$. TESS regularly down-links 2-minute cadence light curves of selected targets and half-hour cadence full-frame images per sector. TESS is optimized to observe cool stars at high precision in order to detect Earth-sized planets. TESS observations of cool stars also capture many stellar flares. In sectors 1 \& 2 alone, 763 flare stars were observed in the 2-minute cadence TESS light curves, with 3247 individual flares recorded \citep{gunter2019}. Cool stars comprise 83\% of these flare stars.

Small flares occur much more frequently than large flares. Although TESS observes each star at high photometric precision for a sufficient amount of time to characterize the occurrence of low-to-moderate energy flares from each cool star, observation times spanning 1-2 sectors are often not long enough to capture the largest superflares. For example, the well-studied flare star Proxima Centauri emits flares of 10$^{32}$ erg or greater on 10 day time scales, but flares of 10$^{33}$ erg or greater on 100 day timescales \citep{Howard2018}. Furthermore, TESS flare observations of each star outside the continuous viewing zone are insensitive to cyclic changes to stellar flaring on timescales longer than 28 days per sector.

The Evryscope \citep{Evryscope2015} is performing long-term high-cadence monitoring of flares and other short-timescale phenomena across the Southern sky, for much longer periods than does TESS. Evryscope is composed of 22 60mm telescopes simultaneously imaging the entire accessible sky at 13$\arcsec$ pixel$^{-1}$. Thus far, Evryscope has produced 2-minute cadence light curves of 15 million sources. While TESS observes each star for $\sim$28 days in the red at high photometric precision, Evryscope observes each star for several years in the blue at moderate precision.

Combining the frequent flares seen in the TESS light curves themselves with rarer Evryscope flares provides for more comprehensive flare monitoring. Evryscope complements TESS by monitoring the high-energy end of each star's flare distribution, as well as any other changes in flare activity that occur on timescales longer than the 28 day observation time per sector. For example, we illustrate in Figure \ref{fig:full_lc} flares in the combined Evryscope and TESS light curves for the case of the active star TIC-231017428 (L 173-39). TESS observed TIC-231017428 for 2 sectors, finding many flares with amplitudes too small to recover with Evryscope, while missing the rarest and largest flares captured by Evryscope.  Future papers in the EvryFlare series will investigate the combined flaring of each star in Evryscope and TESS.

A number of other ground-based surveys are also discovering large flares from nearby stars. High-cadence observations by the Next Generation Transit Survey (NGTS, \citealt{Wheatley2018}) recorded two 10$^{34}$ erg superflares from a bright G8 star \citep{Jackman2018}. NGTS also captured one of the largest M-dwarf flares to be observed to date at high cadence, a 10$^{36.5}$ erg event from a 2 Myr-old M3 star \citep{Jackman2019}. An M-dwarf superflare search using data obtained by the All-Sky Automated Survey for Supernovae (ASAS-SN,\citealt{Shappee2014}) observed 53 large flares, with a bolometric energy range of approximately 10$^{33}$ to 10$^{36}$ erg \citep{Schmidt_2018_ASASSN_flares}. Another M-dwarf flare search in data obtained by the MEarth project \citep{Nutzman2008,Berta2012} discovered 54 large flares from 34 flare stars out of 2226 stars searched \citep{Mondrik2019}. These and other ground-based flare surveys probe different but overlapping regimes: while the high cadence, multi-year observations, and all-sky continuous coverage of Evryscope capture at least an order of magnitude more large flares from early to mid-M dwarfs than other surveys, MEarth and ASAS-SN are best at capturing flares from late M-dwarfs. The ultra-high cadence of NGTS allows unprecedented observations of flare morphology and evolution. The combined flare catalog resulting from all ground-based surveys supplements the 28 days of flare observations TESS provides for each star.

In Section \ref{EvryFlare_Overview} of this work, we describe the Evryscope flare search program, EvryFlare. We describe the Evryscope and its light curve database. We also describe our flare-search sample and our flare search algorithms. In Section \ref{evr_flare_distrib}, we describe our discoveries. These include a number of superflare events that increased the stellar brightness by at least 3 $\textit{g}^{\prime}$ magnitudes, and correlations of flaring with stellar astrophysics. We describe a superflare observed from TOI-455, a nearby TESS Object of Interest that hosts a candidate rocky planet. We describe our constraints on its superflare rate and possible effects to planetary atmospheres. In Section \ref{planet_habitability}, we discuss the implications of extreme flaring for the retention of planetary ozone layers and resulting planetary habitability. In Section \ref{conclude}, we conclude.

\section{The EvryFlare all-sky superflare search}\label{EvryFlare_Overview} \label{EvryFlare}
In order to measure the occurrence of rare superflares, Evryscope monitors the long-term flare activity of all cool stars. We focus the current analysis on bright stars across half the Southern sky. Stellar flares in Evryscope data are discovered and characterized in two independent ways. A brief manual inspection of each Evryscope light curve discovers the largest flares captured by Evryscope. An automated flare search discovers flares of all amplitudes above the photometric noise; these candidate flares are further inspected by eye.

\subsection{Evryscope observations}\label{evryscope_observations} \label{flare_det}
As part of the Evryscope survey of all bright Southern stars, we discover many large stellar flaring events. The Evryscope is an array of small telescopes simultaneously imaging 8150 square degrees and 18,400 square degrees in total each night on the sky at two-minute cadence in \textit{g}\textsuperscript{$\prime$}~\citep{Evryscope2015}. The Evryscope is optimized for bright, nearby stars, with a typical dark-sky limiting magnitude of \textit{g}\textsuperscript{$\prime$}=16. The Evryscope is designed to observe the entire Southern sky down to an airmass of two and at a resolution of 13\arcsec pixel$^{-1}$. To achieve $\sim$6 hours of continuous monitoring each night on each part of the sky, the Evryscope tracks the sky for 2 hours at a time before ratcheting back and continuing observations \citep{Ratzloff2019}.

The Evryscope image archive contains 3.0 million raw images, $\sim$250~TB of data. The Evryscope dataset is reduced at realtime rates by a custom data reduction pipeline \citep{Law2016}. Each image, consisting of a 28.8 MPix FITS file from one camera, is calibrated using a custom wide-field solver. Careful background modeling and subtraction is performed before raw photometry is extracted with forced-apertures at known source positions in a reference catalog. Light curves are then generated for approximately 9.3 million sources across the Southern sky by differential photometry in small sky regions using carefully-selected reference stars and across several apertures \citep{Ratzloff2019}. Any remaining systematics are removed using two iterations of the SysRem detrending algorithm \citep{tamuz2005}.

The Evryscope light curve database is periodically regenerated across the sky for improved photometric precision and longer baseline of observations. The current generation of light curves at the time of this work spans Jan 2016 through June 2018, with an average of 32,000 epochs per star. Light curves of bright stars (\textit{g}\textsuperscript{$\prime$}=10) attain 6 mmag to 1\% photometric precision (depending on the stellar crowding level); light curves of dim stars (\textit{g}\textsuperscript{$\prime$}=15) attain 10\% precision. We note Evryscope precision for dim stars is comparable to TESS precision on dim stars \citep{Ratzloff2019}.

\subsection{Flare search targets}\label{search_sample} 
\begin{figure}[t]
	\centering
    \subfigure
	{
		\includegraphics[trim= -15 10 -80 0, clip, width=3.4in]{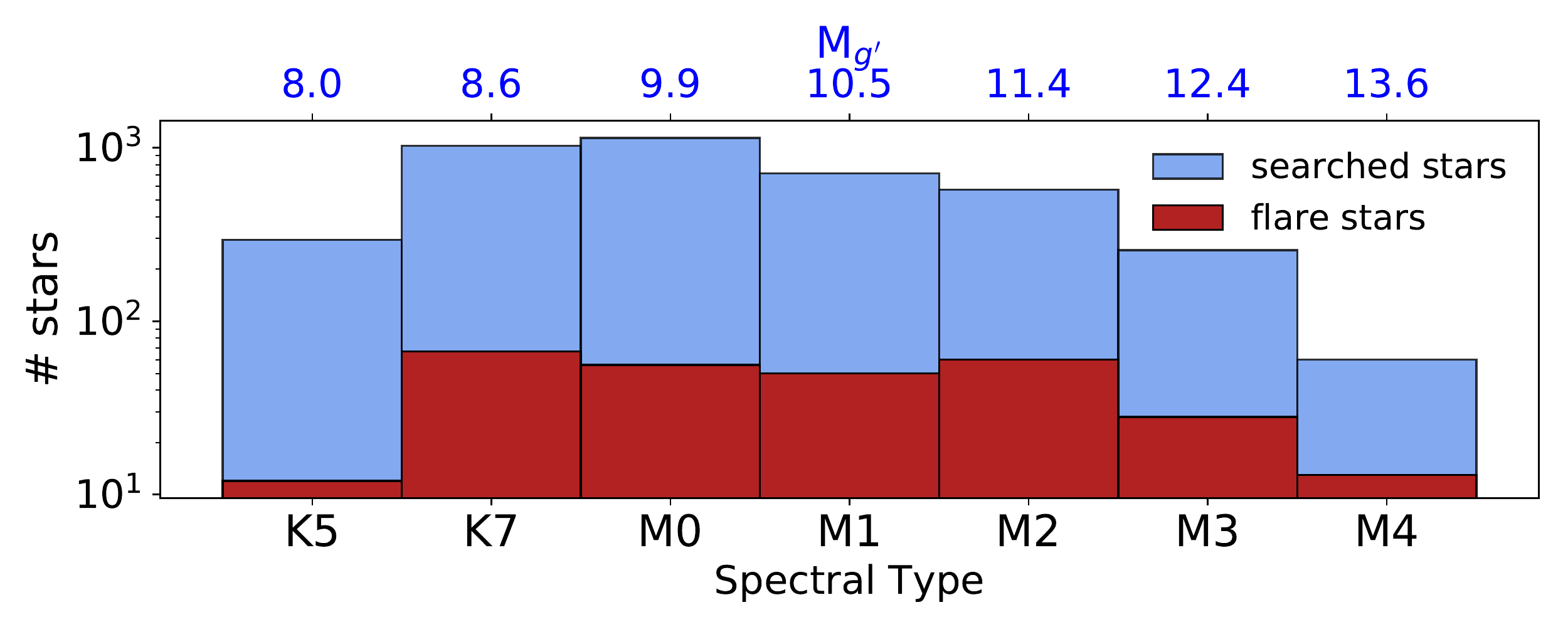}
		\label{fig:indiv_flares_vs_spt}
	}
	\subfigure
	{
		\includegraphics[trim= 10 0 0 12, clip, width=3.4in]{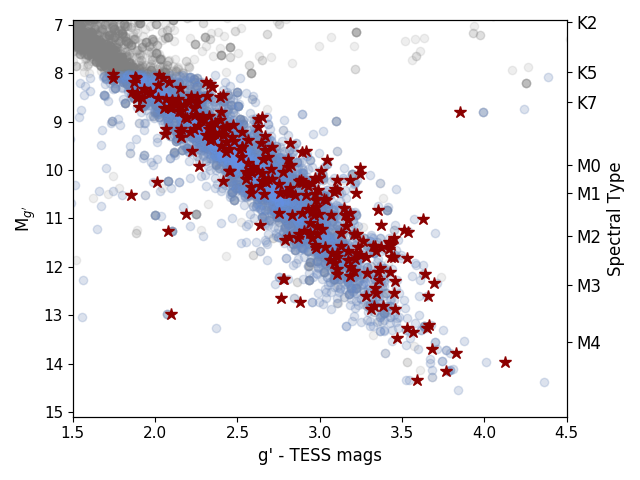}
		\label{fig:flare_stars_vs_spt}
	}
	\caption{Top panel: Evryscope-detected flare stars (red) compared to the full sample of Evryscope light curves of the cool stars (blue). Bottom panel: Our flare-search sample and flare stars plotted on an absolute magnitude versus color diagram. The Evryscope light curves with M$_{g'}>8$ (K5 and later cool stars) are selected in blue and plotted against their Evryscope minus TESS color to ensure main-sequence dwarfs are primarily selected by our simple brightness cut. Evryscope targets earlier than K5 are removed from this analysis of flaring cool stars. The flare stars we observe are plotted as red asterisks. We note the distribution of cool flare stars is slightly offset from the cool star distribution we searched. Because active stars are younger and therefore higher in metallicity than most stars, we expect them to cluster toward the right edge \cite{Mann2015}, as we observe.}
	\label{fig:pop_flares}
\end{figure}

We select cool stars that have both TESS and Evryscope light curves for this subset of the larger EvryFlare search program. We begin with the list of all target stars being observed at two-minute cadence by TESS in  sectors 1 through 6. Due to the large pixel scales of Evryscope and TESS (13" and 21" respectively), we cross-match each target star with Gaia DR2 \citep{Gaia2016,Gaia2018} sources within a 13" aperture. Any star with multiple cross-matches within that radius is discarded if the parallaxes of the cross-matched source differ by more than 1\% or if the distance to the source is greater than 600 pc. The Evryscope forced-photometry catalog is based upon APASS DR9 \citep{Henden2016}; we cross-match each target with its \textit{g}\textsuperscript{$\prime$}-magnitude, discarding any sources without a match.

Using the distance and apparent \textit{g}\textsuperscript{$\prime$}-magnitude, we compute the absolute \textit{g}\textsuperscript{$\prime$}-magnitude and select only targets with $M_{g'}>8$ to constrain our analysis to cool stars with spectral types of $\sim$K5 and later \citep{Kraus2007,Muirhead2018}. We update the J2000 coordinates of high proper motion stars to correct for movement between pixels by J2018 and query the Evryscope light curve database. Of these sources, 20\% do not produce light curves in the Evryscope DB, and 10\% of those remaining are affected by source blending from stellar crowding.

Applying the above constraints, we select 1679 $M_g>8$ Evryscope light curves from a list of 24,816 2-minute cadence targets observed by TESS in Sectors 1 and 2, 1904 Evryscope light curves from a list of 28,577 TESS targets in Sectors 3 and 4, and 1773 Evryscope light curves from a list of 30,840 TESS targets in Sectors 5 and 6. Because some targets are observed in multiple sectors, repeated Evryscope light curves in each list are allowed; we analyze a total of 4068 unique Evryscope light curves.

TESS observes 4212 targets at 2-minute cadence with M$_{g'}>$8, which we flag as likely K5 and later dwarfs. Of these, we analyzed Evryscope light curves for 4068 targets. We exclude earlier-type stars in this flare search. To ensure we are primarily selecting K5 and later stars on the main sequence, we compute the Evryscope g' -magnitude minus TESS magnitude of our sample of light curves. We plot g'-TESS color versus M$_{g'}$ of our final sample of light curves in Figure \ref{fig:pop_flares}. Light curves with Evryscope flares are highlighted.

We search for flares in this cool star subset of the Evryscope light curves. We break up our flare search into sets of Evryscope light curves of two TESS sectors at a time: 1 \& 2, 3 \& 4, and 5 \& 6. The number of epochs (and hence number of flares) in the Evryscope light curves from each batch of sectors will vary by season; seasonal variation of the length of the night is a function of right ascension and therefore TESS Sector. Sectors 1 \& 2 have a median number of 25,134 individual epochs per light curve; sectors 3 \& 4 have a median number of 17,164 epochs per light curve, and sectors 5 \& 6 have a median number of 17,652 epochs per light curve. Stars at the southernmost declinations average $\sim5\times$  these numbers of epochs.

\subsection{Automated search for flares}\label{evr_auto_elfs}
We perform an automated flare search in the Evryscope light curves using a custom flare-search algorithm, Auto-ELFS (Automated Evryscope Light-curve Flare Searcher). Due to the Evryscope ratchet observing length, duration of the night, and various weather interrupts to observing, Auto-ELFS first splits up the light curve into separate ``contiguous" segments of uninterrupted observations, each of which is analyzed separately. Before attempting to locate flares, Auto-ELFS tries to determine which epochs represent the quiescent baseline flux; excluding brightness excursions improves our estimation of the local photometric noise. Likely-quiescent epochs are defined to exclude any epoch that occurs less than 20 minutes following any brightening in magnitude with a significance of 4.5$\sigma$ above an initial estimate of the noise.

Auto-ELFS then searches for flares by applying an exponential-decay matched-filter similar to that of \citet{Liang2016} to the contiguous light curve segment. Peaks in the matched-filter with a filter significance above 4.5$\sigma$ that correspond with peaks in the actual $g^{\prime}$ magnitude light curve with a significance above 2.5$\sigma$ are considered flare candidates. The matched-filter significance is defined as the median-subtracted filter value divided by the standard deviation of the filter values of likely non-flaring epochs. The $g^{\prime}$ light curve significance is defined as the median-subtracted magnitude divided by the standard deviation of the magnitudes of likely non-flaring epochs. We require the flare candidate to be significant in the matched filter in order to recover flares from noisy light curves; we require the flare candidate to be significant in the light curve magnitudes to ensure the flare rises sufficiently above its surrounding epochs to be vetted by eye. Flare start and stop times are determined as the first and last epochs with significance in magnitude (not in filter product) that exceed 1$\sigma$ around the flare peak time.

\begin{figure*}
	\centering
	{
		\includegraphics[trim= 0 0 0 0, clip, width=6.8in]{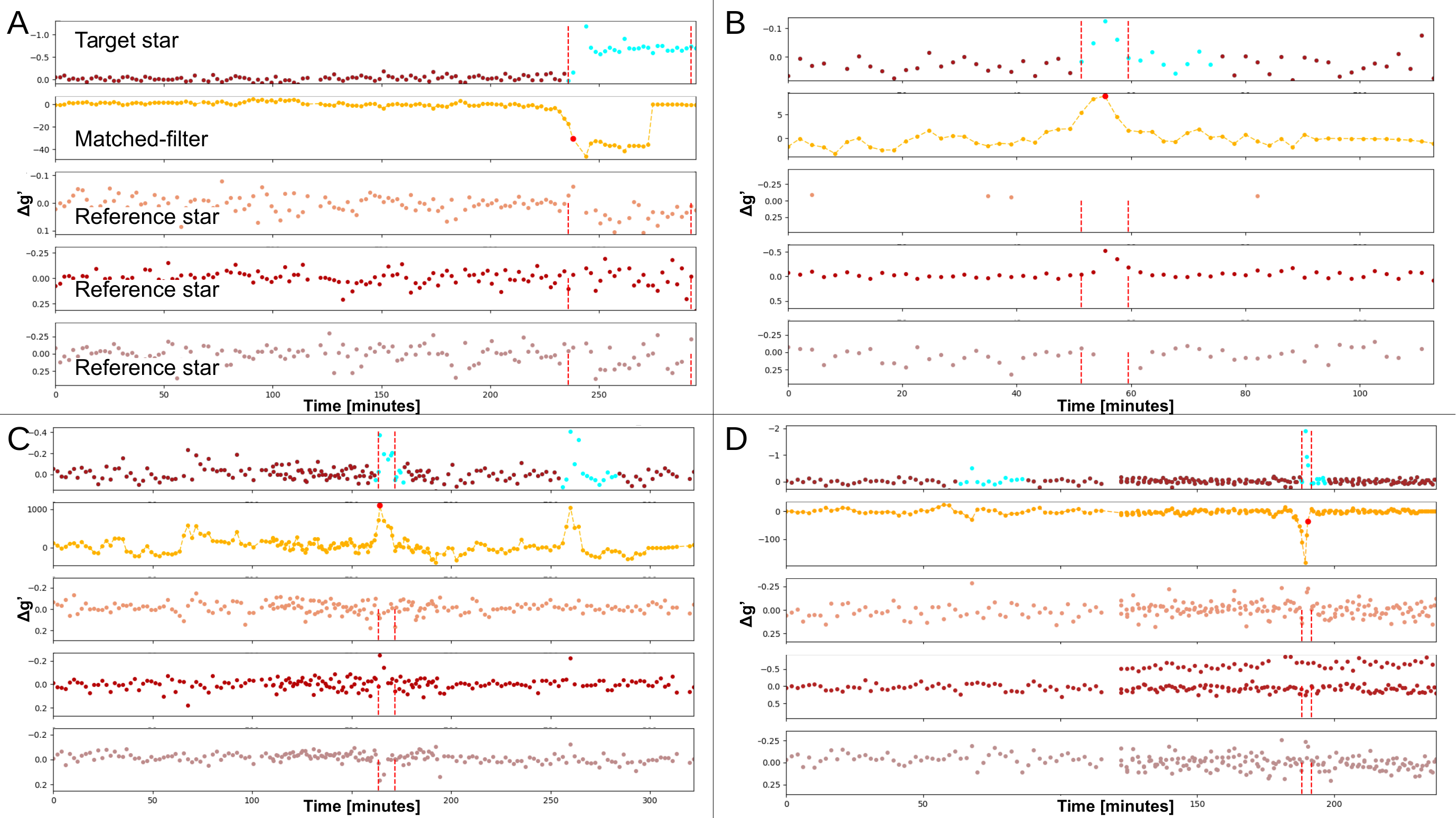}
	}
	\caption{Rejected stellar flare candidates resulting from by-eye vetting. In each vetting image (i.e. A,B,C,D), target star and reference star light curves are displayed, along with the flare matched-filter product. All light curves displayed are in \textit{g}\textsuperscript{$\prime$}, except for the matched-filter (orange) which is unit-less. For each target star, epochs flagged by Auto-ELFS for possible stellar brightening are displayed in blue. Start and stop times for each flare are displayed as vertical dashed red lines. Candidate A is rejected for failing to follow a FRED profile. Candidates B and C are rejected because reference stars display similar behavior. Candidate D is rejected for having an amplitude of multiple magnitudes while lasting less than 10 minutes. This event occurred during telescope shake.}
	\label{fig:FP_byeye_flare_vetting}
\end{figure*}

\begin{figure*}
	\centering
	{
		\includegraphics[trim= 0 0 0 0, clip, width=6.8in]{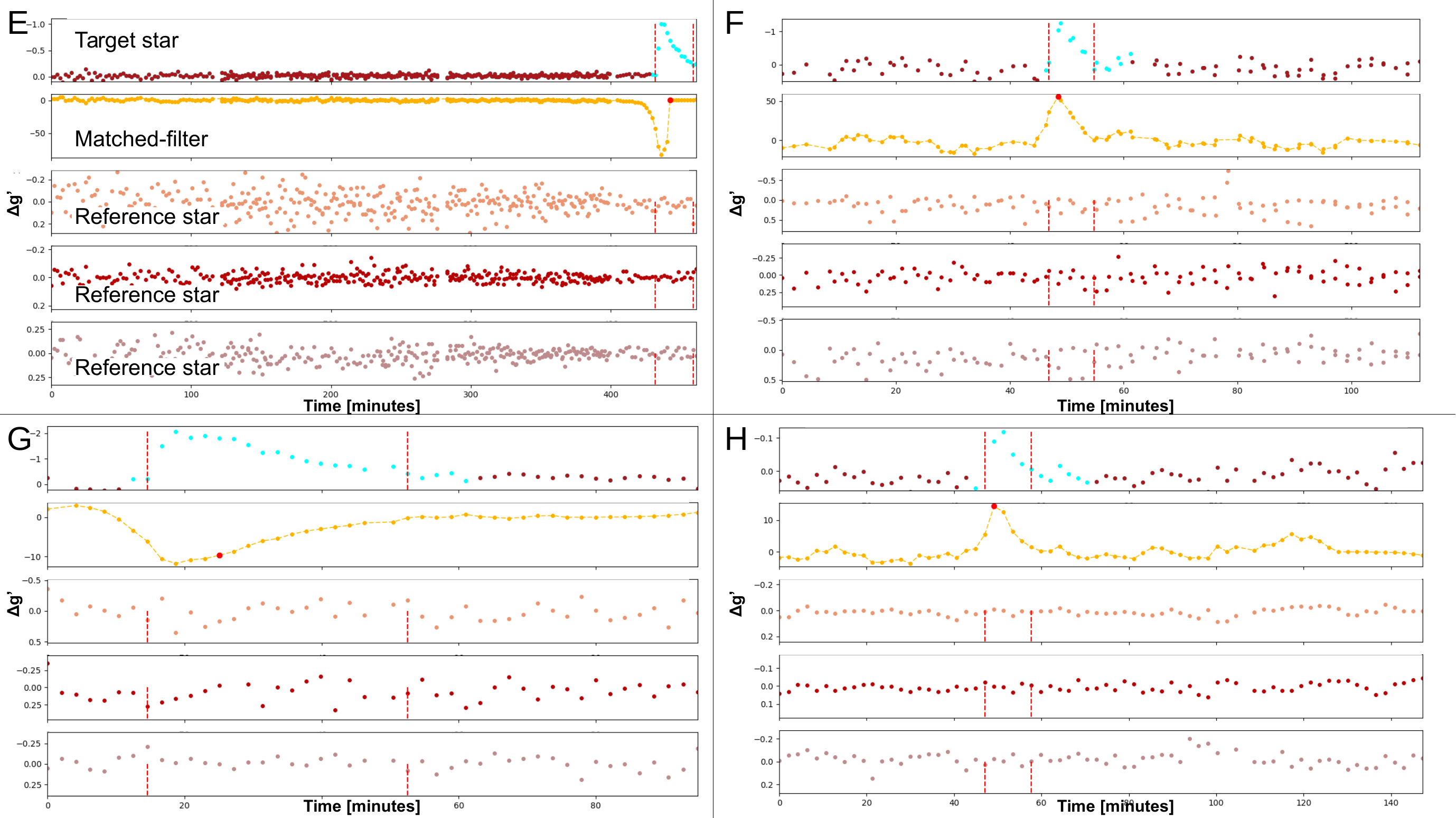}
	}
	\caption{Confirmed stellar flare candidates resulting from by-eye vetting. In each vetting image (i.e. E,F,G,H), target star and reference star light curves are displayed, along with the flare matched-filter product. All light curves displayed are in \textit{g}\textsuperscript{$\prime$}, except for the matched-filter (orange) which is unit-less. For each target star, epochs flagged by Auto-ELFS for possible stellar brightening are displayed in blue. Start and stop times for each flare are displayed as vertical dashed red lines.}
	\label{fig:confirmed_byeye_flare_vetting}
\end{figure*}

Significant candidates are verified by eye in an interactive vetting tool. During interactive vetting, flare candidates from the automated pipeline are confirmed or rejected based on the following criteria: similarity to a FRED profile, dis-similarity to known systematics (such as a Gaussian or box-shaped flare light curve), and a lack of similar flaring behavior at the same time in 3 nearby reference stars. We also exclude from consideration flare candidates that increase in brightness by multiple magnitudes but last less than 10 minutes. Full-frame image cutouts of several of these short multi-magnitude excursions consistently display telescope shake. An example of four flare candidates rejected during vetting for each of these reasons is shown in Figure \ref{fig:FP_byeye_flare_vetting}, and an example of four flares confirmed during vetting are shown in Figure \ref{fig:confirmed_byeye_flare_vetting}.

\subsection{Manual light curve inspection for superflares}\label{evryscope_flare_search} \label{evr_flare_search}
We also perform a brief manual inspection of the entire light curve of each star. Although less sensitive to smaller flares than the automated pipeline, this approach allows us to consistently record the largest flares. Large flares easily observable from the light curve by eye sometimes occur in contiguous segments that last only $\sim$ 20-30 minutes (in periods where due to weather or other observing programs the Evryscope was executing shorter-than-usual ratchets). Auto-ELFS is not designed to operate on contiguous segments of such a short duration due to difficulty in distinguishing in-flare epochs from out-of-flare epochs. Similar difficulties arise when the flare length and the contiguous segment observing length are comparable, e.g. for the largest and longest-lasting superflares, where the slow decay dominates the local background estimation. Light curve inspection remedies this. Finally, some rare systematic brightness excursions of 1-2 magnitude occur consistently across the sky in particular observing seasons but not others. These systematics are readily separated from real flares during manual inspection of all light curves, although they do not occur on the same night for each star and hence do not appear in the 3 nearest reference stars at exactly the same time as the target star. Flares discovered during manual light curve inspection are assigned start and stop times by eye. Flare candidates from this pipeline are subsequently compared against 3 reference stars using the same vetting criteria described in Section \ref{evr_auto_elfs}.

Flares from both automated and manual pipelines are cross-matched against one another and compiled into a single list, keeping one entry for each flare. Because we perform separate searches in each batch of 2 TESS sectors, some of our flare stars will be discovered multiple times. Furthermore, many flares are discovered with an entry from each pipeline. We find 75\% of flares discovered by the manual pipeline are also found by the automated one, and 45\% of flares from the automated pipeline are found in the manual search. Duplicate flares may also occur when long-lasting flares are sometimes ``rediscovered" multiple times by the automated pipeline. Whatever the source of duplicate flare entries, if multiple flare entries are found within 0.1 day of each other, the flare entry with the larger peak magnitude is kept, ensuring the entire flare has been captured and not just the decay tail. This process loses $\sim$2\% of small flares observed near a large flare. Future work will examine the relationships of complex versus single flares occurring in rapid succession after each other \citep{hawley2014,Davenport2016proc} in the Evryscope data-set. Quasi-periodic pulsation (QPP; \citet{Pugh2016}) may be detectable in these complex flares at Evryscope's 2-minute cadence for the brightest flare stars, although most QPPs have  periods and amplitudes below our detection thresholds \citep{McLaughlin2018}.

\subsection{Determination of flare parameters}\label{flare_params}
We describe below how we measure the physical parameters of each individual flare and describe relevant uncertainties:

\begin{itemize}
  \item The \textbf{fractional flux} is calculated as described in \citet{hawley2014}. Fractional flux is computed as $\Delta$F/F=$\frac{\mid F-F_0 \mid}{F_0}$ where $F_0$ is the out-of-flare flux. F$_0$ is determined from the median of the entire light curve in the automated pipeline and from a $\sim$5 day window around the flare in the manual pipeline. 
  \item The \textbf{equivalent duration (ED)} for each flare is calculated as described in \citet{hawley2014}. We compute the ED as ``area-under-the-curve" using the trapezoidal rule, with upper and lower limits of the flare start and stop times. We compute ED as ``area-under-the-curve" rather than as a direct sum of flux received during each 2-minute exposure in order to avoid double-counting flux from flares seen by multiple Evryscope cameras simultaneously. We may safely approximate the ED as ``area-under-the-curve" because the dominant source of error in flare energy is estimation of a star's quiescent energy L$_0$.
  \item We compute the \textbf{quiescent luminosity} in \textit{g}\textsuperscript{$\prime$}~(L$_0$) in erg s$^{-1}$ using the apparent \textit{g}\textsuperscript{$\prime$}~magnitude of the star in the AAVSO Photometric All Sky Survey (APASS) DR9 \citep{Henden2016}, \textit{g}\textsuperscript{$\prime$}~=0 to flux calibration \citep{Hewett2006}, and the Gaia DR2 parallax \citep{Gaia2016,Gaia2018}.
  \item Flare \textbf{energy in the Evryscope \textit{g}\textsuperscript{$\prime$}~bandpass} is given in erg by ED$\times$L$_0$. 
  \item We convert the flare energy in the Evryscope bandpass into \textbf{bolometric energy} using the energy partitions of \citet{Osten2015}. We estimate the bolometric flare energy of a 9000 K flare blackbody with emission matching the measured Evryscope flux; the fraction of the bolometric energy found in the Evryscope \textit{g}\textsuperscript{$\prime$}  bandpass is $f_{g'}$=0.19.
  \item The \textbf{full-width-at-half-maximum} (FWHM, in minutes) of each flare was recorded by an automated algorithm to estimate the distribution of highly impulsive flares as described in \citet{Kowalski2013}. As such, we estimate the FWHM as 2 minutes of rise/decay time plus the elapsed time between the first and last points at or above 50\% of the peak flare flux. We compared the FWHM computed this way versus a FWHM computed as 2 minutes times the number of points above the 50\% flux and found both values agreed for dozens of flares, but only when one camera recorded each flare. The number-of-points method doubled the FWHM when the flare was observed by 2 overlapping cameras.
  \item The \textbf{impulse} of each flare was then recorded as the flare peak fractional flux divided by the FWHM in minutes.
\end{itemize}

These values and relevant uncertainties are recorded for each flare in Table \ref{table:indiv_flares_tab}. We here summarize the errors of our flare parameters (units same as in Table \ref{table:indiv_flares_tab}): Uncertainties in peak flare time and FWHM result from the observing cadence and should average $\sim$2 minutes. Uncertainties in ED and flare energy are computed as the inverse significance of detection; these uncertainties are computed at an average $\sim$10\% error because the median and 1$\sigma$ spread in significance of detection is 10.2$^{7}_{-4}$. Errors in flare amplitude are computed as the photometric errors at the peak flare times. In $\Delta g^{\prime}$, the median and spread of the errors is given by 0.02$^{0.03}_{-0.01}$, and in fractional-flux, the median and spread of the errors is given by 0.01$^{0.05}_{-0.009}$. The median and spread of the errors in flare impulse is given by 0.05$^{0.05}_{-0.3}$. Errors in L$_0$ and M$_{g^{\prime}}$ both depend only on Gaia DR2 parallaxes and APASS DR9 g-magnitudes, which both have typical errors below the 10\% level. Photometric spectral types estimated from M$_{g'}$ are approximate, and are accurate within 1-2 spectral sub-types.

\begin{table*}
\caption{Individual Flares Observed by Evryscope from Cool Stars in TESS Sectors 1-6}
\begin{tabular}{p{1.5cm} p{1.2cm} p{1.5cm} p{1.1cm} p{1.1cm} p{1.1cm} p{1.1cm} p{0.9cm} p{1.0cm} p{1.3cm} p{0.9cm} p{0.9cm}}
\hline
&  &  &  &  &  &  &  &  &  &  &  \\
TIC ID & Sector & Flare time & log\newline E$_\mathrm{bol}$ & Contrast & Peak flux & log L$_0$ & ED & FWHM & Impulse & M$_{\textit{g}\textsuperscript{$\prime$}}$ & SpT \\
 & & [MJD] & [log erg] & [$\Delta$M$_{\textit{g}\textsuperscript{$\prime$}}$] & [$\Delta$F/F] & [log \newline erg s$^{-1}$] & [s] & [min] & [$\Delta$F/F/min] & & \\
\hline
&  &  &  &  &  &  &  &  &  &  &  \\
 &  &  &  &  & ... &  &  &  &  &  &  \\
 &  &  &  &  &  &  &  &  &  &  &  \\
348839788 & 1-6 & 57835.110 & 33.5 & 0.2 & 0.2 & 31.0 & 60 & 6.4 & 0.03 & 9.9 & M0 \\
177309077 & 1,4,6 & 57477.182 & 34.8 & 0.9 & 1.4 & 31.4 & 420 & 6.2 & 0.22 & 8.9 & K7 \\
304032310 & 1 & 57416.122 & 33.9 & 1.0 & 1.6 & 30.3 & 830 & 6.3 & 0.26 & 11.8 & M2 \\
232077453 & 1,2 & 57663.100 & 34.8 & 2.8 & 12.4 & 30.8 & 1810 & 2.0 & 6.18 & 10.5 & M1 \\
348763552 & 1 & 57610.186 & 33.7 & 1.2 & 2.0 & 30.5 & 270 & 2.0 & 1.0 & 11.2 & M2 \\
219315573 & 1 & 57697.044 & 33.5 & 1.9 & 4.8 & 29.8 & 780 & 2.4 & 2.0 & 12.9 & M3 \\
220432563 & 1,2,3,5,6 & 57601.322 & 34.1 & 1.8 & 4.4 & 30.2 & 1480 & 4.1 & 1.09 & 12.0 & M3 \\
220432563 & 1,2,3,5,6 & 57609.389 & 33.3 & 0.5 & 0.7 & 30.2 & 260 & 8.2 & 0.08 & 12.0 & M3 \\
220432563 & 1,2,3,5,6 & 57610.430 & 33.5 & 1.2 & 1.9 & 30.2 & 440 & 6.1 & 0.31 & 12.0 & M3 \\
220432563 & 1,2,3,5,6 & 57699.096 & 33.7 & 1.0 & 1.6 & 30.2 & 590 & 4.1 & 0.4 & 12.0 & M3 \\
220432563 & 1,2,3,5,6 & 57755.160 & 33.4 & 0.7 & 0.9 & 30.2 & 340 & 4.2 & 0.23 & 12.0 & M3 \\
220432563 & 1,2,3,5,6 & 58051.190 & 34.0 & 1.6 & 3.5 & 30.2 & 1210 & 4.1 & 0.86 & 12.0 & M3 \\
220432563 & 1,2,3,5,6 & 58108.170 & 34.4 & 2.3 & 7.4 & 30.2 & 3000 & 6.1 & 1.21 & 12.0 & M3 \\
220432563 & 1,2,3,5,6 & 58195.033 & 34.2 & 1.8 & 4.8 & 30.2 & 2090 & 2.0 & 2.39 & 12.0 & M3 \\
355767588 & 1,2 & 57955.210 & 34.5 & 0.6 & 0.6 & 31.5 & 200 & 4.1 & 0.15 & 8.7 & K7 \\
150185574 & 2,3,4,5,6 & 57771.237 & 32.7 & 0.2 & 0.2 & 30.8 & 20 & 2.0 & 0.11 & 10.5 & M1 \\
316805931 & 2 & 57662.008 & 35.0 & 1.1 & 1.6 & 31.4 & 720 & 6.1 & 0.27 & 8.9 & K7 \\
120606992 & 2 & 57682.207 & 34.2 & 0.9 & 1.2 & 31.3 & 160 & 2.0 & 0.62 & 9.2 & K7 \\
5611068 & 2 & 58066.056 & 34.7 & 0.2 & 0.2 & 31.5 & 340 & 30.6 & 0.01 & 8.8 & K7 \\
5796048 & 2 & 57923.345 & 34.2 & 1.1 & 1.9 & 30.8 & 440 & 4.1 & 0.47 & 10.5 & M1 \\
369707376 & 1,2,3 & 57634.211 & 32.8 & 0.3 & 0.3 & 30.1 & 90 & 2.0 & 0.14 & 12.1 & M3 \\
369707376 & 1,2,3 & 57637.239 & 32.4 & 0.4 & 0.4 & 30.1 & 30 & 2.0 & 0.2 & 12.1 & M3 \\
394137224 & 1 & 57609.340 & 33.1 & 0.4 & 0.4 & 30.8 & 40 & 2.3 & 0.17 & 10.4 & M1 \\
394137224 & 1 & 57696.060 & 33.6 & 0.4 & 0.4 & 30.8 & 100 & 4.4 & 0.09 & 10.4 & M1 \\
237885807 & 1,2 & 57697.030 & 34.1 & 1.6 & 3.3 & 30.4 & 870 & 4.1 & 0.83 & 11.5 & M2 \\
237885807 & 1,2 & 57982.277 & 34.0 & 1.1 & 1.7 & 30.4 & 740 & 6.6 & 0.26 & 11.5 & M2 \\
381949148 & 1-6 & 58020.357 & 34.7 & 0.3 & 0.4 & 31.3 & 520 & 41.0 & 0.01 & 9.3 & M0 \\
381949148 & 1-6 & 58105.136 & 34.1 & 0.3 & 0.3 & 31.3 & 120 & 10.2 & 0.03 & 9.3 & M0 \\
207199350 & 2,3,4 & 58143.177 & 34.2 & 1.0 & 1.6 & 30.6 & 700 & 6.4 & 0.25 & 10.9 & M1 \\
300907829 & 1 & 58292.398 & 34.5 & 0.8 & 1.1 & 30.9 & 810 & 12.7 & 0.08 & 10.2 & M1 \\
149914329 & 2,3,4,6 & 57476.018 & 34.8 & 1.4 & 2.7 & 31.3 & 590 & 4.1 & 0.66 & 9.2 & K7 \\
149914329 & 2,3,4,6 & 57642.356 & 33.3 & 0.5 & 0.5 & 31.3 & 20 & 2.0 & 0.26 & 9.2 & K7 \\
149914329 & 2,3,4,6 & 57753.071 & 34.2 & 1.0 & 1.4 & 31.3 & 150 & 4.1 & 0.34 & 9.2 & K7 \\
149914329 & 2,3,4,6 & 57755.152 & 34.0 & 0.4 & 0.5 & 31.3 & 100 & 7.8 & 0.06 & 9.2 & K7 \\
152877086 & 1 & 57631.253 & 33.7 & 0.5 & 0.7 & 30.5 & 320 & 8.5 & 0.08 & 11.3 & M2 \\
471016669 & 2 & 57710.128 & 33.3 & 0.5 & 0.6 & 30.1 & 310 & 8.5 & 0.07 & 12.3 & M3 \\
471016669 & 2 & 58069.094 & 32.4 & 0.3 & 0.3 & 30.1 & 40 & 4.1 & 0.06 & 12.3 & M3 \\
350223741 & 1 & 58096.096 & 35.1 & 2.8 & 13.4 & 30.3 & 9590 & 6.8 & 1.96 & 11.6 & M2 \\
266998480 & 1 & 57623.238 & 33.7 & 0.4 & 0.4 & 31.1 & 70 & 6.4 & 0.06 & 9.7 & M0 \\
270298604 & 1 & 58291.271 & 34.1 & 0.7 & 0.9 & 30.4 & 920 & 22.7 & 0.04 & 11.4 & M2 \\
175490502 & 2 & 58249.40 & 34.1 & 1.7 & 3.6 & 30.6 & 580 & 2.0 & 1.79 & 10.9 & M1 \\
308453663 & 1,2,4,5 & 57839.112 & 33.7 & 0.4 & 0.4 & 31.6 & 30 & 2.0 & 0.21 & 8.5 & K7 \\
201248233 & 1,2 & 57642.319 & 33.7 & 0.4 & 0.4 & 30.7 & 180 & 2.0 & 0.22 & 10.8 & M1 \\
302965929 & 1,5 & 57682.024 & 34.3 & 0.3 & 0.3 & 31.3 & 170 & 13.0 & 0.03 & 9.1 & K7 \\
302965929 & 1,5 & 58140.305 & 35.2 & 1.3 & 2.3 & 31.3 & 1510 & 4.2 & 0.55 & 9.1 & K7 \\
111184885 & 2 & 57636.083 & 33.6 & 0.1 & 0.1 & 31.6 & 20 & 2.0 & 0.06 & 8.5 & K7 \\
220433363 & 2,3,4,5,6 & 57401.227 & 33.9 & 0.6 & 0.8 & 30.5 & 510 & 8.4 & 0.1 & 11.3 & M2 \\
220433363 & 2,3,4,5,6 & 57420.134 & 32.7 & 0.1 & 0.1 & 30.5 & 30 & 4.1 & 0.03 & 11.3 & M2 \\
220433363 & 2,3,4,5,6 & 57421.217 & 33.1 & 0.1 & 0.1 & 30.5 & 80 & 10.5 & 0.01 & 11.3 & M2 \\
220433363 & 2,3,4,5,6 & 57445.105 & 33.8 & 0.7 & 0.9 & 30.5 & 370 & 4.2 & 0.21 & 11.3 & M2 \\
220433363 & 2,3,4,5,6 & 57663.289 & 33.5 & 0.3 & 0.3 & 30.5 & 180 & 4.1 & 0.08 & 11.3 & M2 \\
220433363 & 2,3,4,5,6 & 57671.329 & 34.8 & 0.9 & 1.4 & 30.5 & 3800 & 46.9 & 0.03 & 11.3 & M2 \\
220433363 & 2,3,4,5,6 & 57683.303 & 33.3 & 0.3 & 0.4 & 30.5 & 130 & 6.1 & 0.06 & 11.3 & M2 \\
220433363 & 2,3,4,5,6 & 57695.154 & 32.8 & 0.2 & 0.2 & 30.5 & 40 & 2.0 & 0.09 & 11.3 & M2 \\
220433363 & 2,3,4,5,6 & 57750.181 & 33.1 & 0.3 & 0.3 & 30.5 & 80 & 4.2 & 0.08 & 11.3 & M2 \\
220433363 & 2,3,4,5,6 & 57992.390 & 33.4 & 0.2 & 0.2 & 30.5 & 140 & 6.1 & 0.04 & 11.3 & M2 \\
220433363 & 2,3,4,5,6 & 58019.160 & 33.2 & 0.2 & 0.2 & 30.5 & 90 & 8.2 & 0.02 & 11.3 & M2 \\
220433363 & 2,3,4,5,6 & 58069.228 & 33.5 & 0.4 & 0.5 & 30.5 & 180 & 6.1 & 0.07 & 11.3 & M2 \\
220433363 & 2,3,4,5,6 & 58075.057 & 33.2 & 0.4 & 0.4 & 30.5 & 100 & 2.0 & 0.2 & 11.3 & M2 \\
220433363 & 2,3,4,5,6 & 58182.049 & 33.3 & 0.5 & 0.6 & 30.5 & 130 & 2.0 & 0.32 & 11.3 & M2 \\
 &  &  &  &  & ... &  &  &  &  &  &  \\
 &  &  &  &  &  &  &  &  &  &  &  \\
\hline
\end{tabular}
\label{table:indiv_flares_tab}
{\newline\newline \textbf{Notes.} The parameters of 575 individual large flares recorded by Evryscope (1 flare per row). This is a subset of the full table. The full table is available in machine-readable form, with uncertainties to parameters where applicable in addition to the columns displayed here. The columns here are: TIC ID, the TESS sector(s), the flare event time in MJD, the bolometric energy of the flare in log erg, the flare amplitude (contrast) in \textit{g}$^{\prime}$ magnitudes, the flare amplitude in fractional flux units, the stellar quiescent luminosity in \textit{g}$^{\prime}$ in log erg/sec, the equivalent duration in \textit{g}$^{\prime}$ in sec, the flare FWHM in minutes, the flare impulse defined as peak fractional flux / FWHM in minutes, the stellar absolute magnitude in \textit{g}$^{\prime}$, and the spectral type estimated from M$_{g^{\prime}}$.}
\end{table*}

\begin{turnpage}

\begin{table*}

\caption{Cool Flare Stars Observed by Evryscope in TESS Sectors 1-6}
\begin{tabular}{p{1.4cm} p{1.2cm} p{1.5cm} p{1.0cm} p{0.8cm} p{0.8cm} p{0.9cm} p{0.7cm} p{0.7cm} p{0.8cm} p{0.8cm} p{0.8cm} p{0.8cm} p{0.9cm} p{0.7cm} p{0.7cm} p{1.1cm} p{0.6cm} p{0.6cm}}
\hline
 &  &  &  &  &  &  &  &  &  &  &  &  &  &  &  &  &  &  \\
 & & & & & & & & & Mean & Max & Mean & Max & Gal. & & & & & \\
TIC ID & RA & Dec & Sector & Flares & Cont.\newline Obs. & Super \newline flares &  $\alpha_\mathrm{FFD}$ & $\beta_\mathrm{FFD}$ & log \newline E$_\mathrm{bol}$ & log \newline E$_\mathrm{bol}$ & Contr. & Contr. & Lat. & T & \textit{g}\textsuperscript{$\prime$} & log L$_0$ & M$_{\textit{g}\textsuperscript{$\prime$}}$ & SpT \\
& [J2018] & [J2018] & & & [d] & [yr$^{-1}$] & & & [log\newline erg] & [log \newline erg] & [$\Delta$M$_{\textit{g}\textsuperscript{$\prime$}}$] & [$\Delta$M$_{\textit{g}\textsuperscript{$\prime$}}$] & [deg.] & & & [log \newline erg s$^{-1}$] & & \\
\hline
 &  &  &  &  &  &  &  &  &  &  &  &  &  &  &  &  &  &  \\
 &  &  &  &  &  &  &  & ... &  &  &  &  &  &  &  &  &  &  \\
 &  &  &  &  &  &  &  &  &  &  &  &  &  &  &  &  &  &  \\
593228 & 71.84526 & -27.84147 & 5 & 1 & 24.2 & 15.1 &  &  & 34.6 & 34.6 & 1.1 & 1.1 & -38.2 & 9.3 & 10.7 & 31.0 & 9.9 & M0 \\
593230 & 71.84526 & -27.84147 & 5 & 1 & 24.2 & 15.1 &  &  & 34.6 & 34.6 & 1.1 & 1.1 & -38.2 & 8.0 & 10.7 & 31.4 & 8.9 & K7 \\
671393 & 73.40664 & -28.59176 & 5 & 1 & 15.8 & 23.1 &  &  & 50.8 & 50.8 & 2.6 & 2.6 & -37.05 & 12.0 & 14.7 & 30.3 & 11.8 & M2 \\
1273249 & 72.67425 & -31.27464 & 5 & 3 & 25.5 & 42.9 &  &  & 34.9 & 35.1 & 0.5 & 0.6 & -38.26 & 10.0 & 12.3 & 30.8 & 10.5 & M1 \\
5611068 & 346.03061 & -17.13679 & 2 & 1 & 23.0 & 15.9 &  &  & 34.7 & 34.7 & 0.2 & 0.2 & -63.29 & 13.4 & 11.7 & 30.5 & 11.2 & M2 \\
5656273 & 347.0817 & -15.4098 & 2 & 5 & 22.5 & 113.8 & -0.33 & 10.4 & 35.0 & 35.7 & 0.7 & 2.1 & -63.35 & 9.2 & 11.6 & 29.8 & 12.9 & M3 \\
5796048 & 348.9343 & -12.36468 & 2 & 1 & 21.7 & 16.8 &  &  & 34.2 & 34.2 & 1.1 & 1.1 & -63.16 & 10.6 & 13.5 & 30.2 & 12.0 & M3 \\
7625199 & 64.41912 & -37.77351 & 4,5 & 1 & 35.4 & 10.3 &  &  & 34.5 & 34.5 & 0.4 & 0.4 & -45.78 & 11.2 & 13.5 & 30.2 & 12.0 & M3 \\
9210746 & 352.55716 & -20.39162 & 2 & 4 & 21.0 & 69.5 &  &  & 33.9 & 34.3 & 0.3 & 0.4 & -70.27 & 8.7 & 11.8 & 30.2 & 12.0 & M3 \\
10863087 & 25.93832 & -6.04458 & 3 & 1 & 25.3 & 14.4 &  &  & 33.1 & 33.1 & 0.6 & 0.6 & -65.52 & 10.4 & 13.8 & 30.2 & 12.0 & M3 \\
21627442 & 87.12652 & -41.45889 & 6 & 1 & 42.9 & 8.5 &  &  & 35.1 & 35.1 & 0.7 & 0.7 & -28.87 & 10.2 & 12.3 & 30.2 & 12.0 & M3 \\
24452802 & 82.43626 & -32.6539 & 5,6 & 1 & 17.0 & 21.5 &  &  & 34.4 & 34.4 & 1.7 & 1.7 & -30.5 & 10.9 & 14.5 & 30.2 & 12.0 & M3 \\
24662390 & 80.92147 & -8.28237 & 5 & 1 & 5.8 & 62.9 &  &  & 34.0 & 34.0 & 0.1 & 0.1 & -23.25 & 11.5 & 14.0 & 30.2 & 12.0 & M3 \\
29853348 & 23.80831 & -7.21444 & 3 & 4 & 18.3 & 79.8 &  &  & 34.6 & 35.1 & 1.4 & 2.1 & -67.54 & 10.6 & 14.1 & 30.2 & 12.0 & M3 \\
29919288 & 24.92133 & -7.03761 & 3 & 1 & 11.3 & 32.3 &  &  & 33.8 & 33.8 & 1.7 & 1.7 & -66.87 & 11.4 & 14.7 & 31.5 & 8.7 & K7 \\
31740375 & 48.31711 & -66.4077 & 1,2,3 & 2 & 19.0 & 38.4 &  &  & 33.8 & 34.0 & 1.5 & 1.6 & -45.07 & 11.8 & 14.9 & 30.8 & 10.5 & M1 \\
32811633 & 86.45644 & -27.16056 & 6 & 1 & 28.4 & 12.9 &  &  & 34.2 & 34.2 & 0.8 & 0.8 & -25.58 & 11.8 & 12.8 & 31.4 & 8.9 & K7 \\
33839863 & 64.71625 & -72.75114 & 2,3,4,5,6 & 1 & 44.9 & 8.1 &  &  & 35.5 & 35.5 & 2.4 & 2.4 & -36.76 & 12.2 & 14.5 & 31.3 & 9.2 & K7 \\
33864387 & 65.41142 & -72.56545 & 2,3,4,5,6 & 2 & 54.7 & 13.3 &  &  & 34.0 & 34.1 & 0.7 & 0.8 & -36.67 & 11.4 & 14.4 & 31.5 & 8.8 & K7 \\
38461205 & 61.24094 & -63.90868 & 1,2,3,4,5 & 1 & 37.3 & 9.8 &  &  & 35.0 & 35.0 & 1.9 & 1.9 & -42.03 & 11.7 & 14.4 & 30.8 & 10.5 & M1 \\
38583513 & 63.52203 & -65.41773 & 1-6 & 1 & 59.6 & 6.1 &  &  & 34.2 & 34.2 & 0.6 & 0.6 & -40.49 & 14.1 & 14.0 & 30.1 & 12.1 & M3 \\
38586438 & 64.08964 & -62.0132 & 1,2,4,5,6 & 1 & 35.4 & 10.3 &  &  & 33.9 & 33.9 & 0.8 & 0.8 & -41.62 & 11.6 & 14.6 & 30.1 & 12.1 & M3 \\
38631589 & 57.68297 & -6.10284 & 5 & 1 & 29.4 & 0.0 &  &  & 32.7 & 32.7 & 0.9 & 0.9 & -42.54 & 13.2 & 13.6 & 30.8 & 10.4 & M1 \\
38814531 & 67.98366 & -61.03467 & 1-6 & 1 & 47.3 & 7.7 &  &  & 34.9 & 34.9 & 2.3 & 2.3 & -40.2 & 11.8 & 14.4 & 30.8 & 10.4 & M1 \\
43605290 & 75.48672 & -6.94835 & 5 & 1 & 26.3 & 0.0 &  &  & 32.5 & 32.5 & 0.9 & 0.9 & -27.47 & 9.4 & 13.0 & 30.4 & 11.5 & M2 \\
44796808 & 62.41391 & -26.8143 & 4,5 & 1 & 20.4 & 17.9 &  &  & 33.7 & 33.7 & 0.6 & 0.6 & -46.13 & 10.8 & 13.7 & 30.4 & 11.5 & M2 \\
49593799 & 37.03401 & -36.47239 & 3 & 1 & 10.1 & 36.1 &  &  & 33.7 & 33.7 & 1.9 & 1.9 & -67.51 & 11.3 & 14.8 & 31.3 & 9.3 & M0 \\
49672084 & 95.67231 & -27.63081 & 6 & 2 & 17.3 & 42.2 &  &  & 33.8 & 33.9 & 1.2 & 1.5 & -18.07 & 11.0 & 14.3 & 31.3 & 9.3 & M0 \\
50436541 & 35.54551 & -76.80395 & 1,2 & 3 & 77.5 & 14.1 &  &  & 34.3 & 34.4 & 0.3 & 0.4 & -39.15 & 10.2 & 11.9 & 30.6 & 10.9 & M1 \\
50745582 & 83.01872 & -3.0913 & 6 & 2 & 14.0 & 52.1 &  &  & 34.0 & 34.1 & 0.3 & 0.3 & -19.04 & 9.3 & 12.1 & 30.9 & 10.2 & M1 \\
55368621 & 77.4463 & -60.00157 & 1-6 & 1 & 60.4 & 6.0 &  &  & 34.2 & 34.2 & 0.4 & 0.4 & -35.86 & 11.6 & 13.8 & 31.3 & 9.2 & K7 \\
55604374 & 73.65541 & -60.23338 & 1-6 & 1 & 53.0 & 6.9 &  &  & 34.7 & 34.7 & 0.7 & 0.7 & -37.69 & 11.8 & 14.3 & 31.3 & 9.2 & K7 \\
64053930 & 38.49955 & -18.19797 & 4 & 1 & 12.8 & 28.5 &  &  & 35.3 & 35.3 & 1.7 & 1.7 & -64.77 & 11.5 & 14.7 & 31.3 & 9.2 & K7 \\
66499720 & 356.23364 & -18.47261 & 2 & 1 & 16.0 & 22.8 &  &  & 34.2 & 34.2 & 1.0 & 1.0 & -72.42 & 11.8 & 14.4 & 31.3 & 9.2 & K7 \\
69889554 & 338.65616 & -19.60883 & 2 & 1 & 21.5 & 17.0 &  &  & 34.4 & 34.4 & 0.9 & 0.9 & -57.82 & 11.5 & 14.2 & 30.5 & 11.3 & M2 \\
 &  &  &  &  &  &  &  & ... &  &  &  &  &  &  &  &  &  &  \\
 &  &  &  &  &  &  &  &  &  &  &  &  &  &  &  &  &  &  \\
\hline
\end{tabular}
\label{table:flare_stars_tab}
{\newline\newline \textbf{Notes.} Parameters of 284 flare stars monitored by Evryscope (1 star per row). This is a subset of the full table. The full table is available in machine-readable form, with uncertainties to parameters where applicable in addition to the columns displayed here. The columns here are: TIC ID, RA and Dec (the current Evryscope-measured positions of the star), the TESS sector(s) the star was observed, the number of flares observed by Evryscope, the total continuous Evryscope observation time in days obtained over 2 years, the annual superflare rate, the FFD parameter $\alpha$, the FFD parameter $\beta$, the mean flare energy in log erg, the maximum flare energy observed in log erg, the mean flare amplitude (contrast) in $\Delta$\textit{g}$^{\prime}$ magnitudes, the maximum flare amplitude (contrast) observed in $\Delta$\textit{g}$^{\prime}$ magnitudes, the galactic latitude in degrees, the stellar TESS-magnitude, the stellar \textit{g}$^{\prime}$ magnitude, the stellar quiescent luminosity in \textit{g}$^{\prime}$ in log erg/sec, the stellar absolute magnitude in \textit{g}$^{\prime}$, and the spectral type estimated from M$_{g^{\prime}}$.}
\end{table*}

\end{turnpage}

\subsection{Flare frequency distributions}\label{evry_flare}
To estimate the superflare rate for each star, the number of flares observed and the total observing time are calculated. We compute the total observing time as the number of epochs in each light curve times a two-minute exposure. We ignore the effect of double-counting epochs from occasional camera overlaps on the total observing time, as only $\sim$10\% of epochs are doubled and the observing time is not the dominant source of error.

For stars with less than five flares, we estimate the superflare rate as the number of superflares actually observed divided by the total observing time. Limits on non-flaring stars are large; we focus this work upon stars with at least one flare observed. The upper and lower limits on the superflare rate are given by a 1$\sigma$ binomial confidence interval.

For stars with at least five flares, we calculate the cumulative flare frequency distribution (FFD) by fitting a cumulative power-law to the flares, and estimating the uncertainty in our fit through 1000 Monte-Carlo posterior draws consistent with our uncertainties in occurrence rates. We represent the cumulative FFD in bolometric energy by a power law of the form $\log{\nu}=\alpha\log{E} + \beta$, where $\nu$ is the number of flares with an energy greater than or equal to $E$ erg per day, $\alpha$ gives the frequency at which flares of various energies occur, and $\beta$ is the y-intercept and sets the overall rate of flaring. We calculate the uncertainty in the cumulative occurrence for each Evryscope flare with a binomial $1\sigma$ confidence interval statistic (following \citealt{davenport2016}). The observation time, number of flares observed, estimated $\alpha$, $\beta$, superflare rates, and uncertainties on these parameters are recorded in Table \ref{table:flare_stars_tab}. Following \citet{gunter2019}, we also include in Table \ref{table:flare_stars_tab} the maximum and mean amplitude and bolometric energy of each Evryscope flare star for comparison. Because we are observing a large sample of large flares, we compute the FFD of each star without weighting recovery completeness using flare injection-and-recovery.

\begin{figure}[t]
	\centering
	\subfigure
	{
		\includegraphics[trim= 1 0 0 12, clip, width=3.4in]{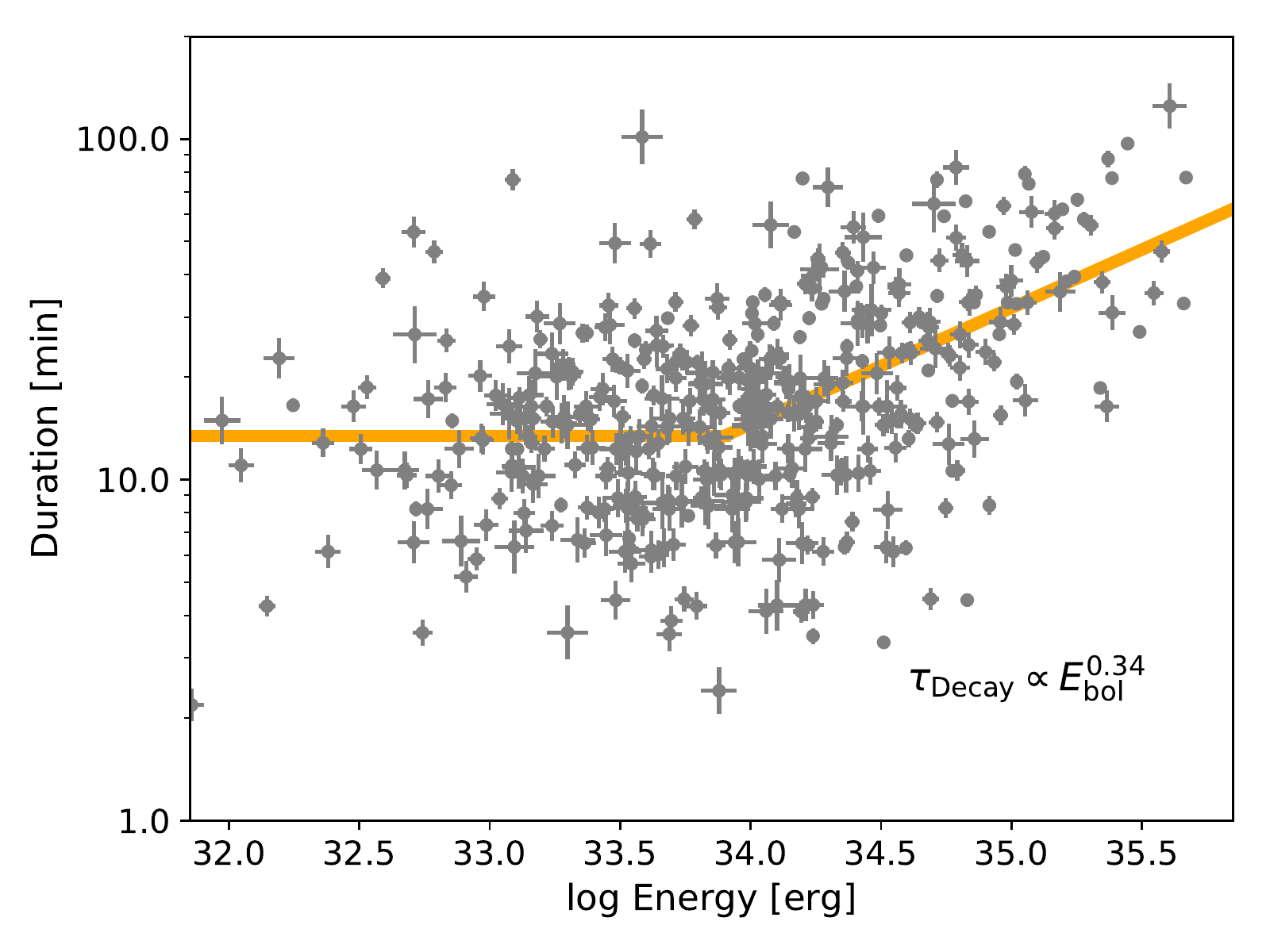}
	}
	{
		\includegraphics[trim= 1 0 0 12, clip, width=3.4in]{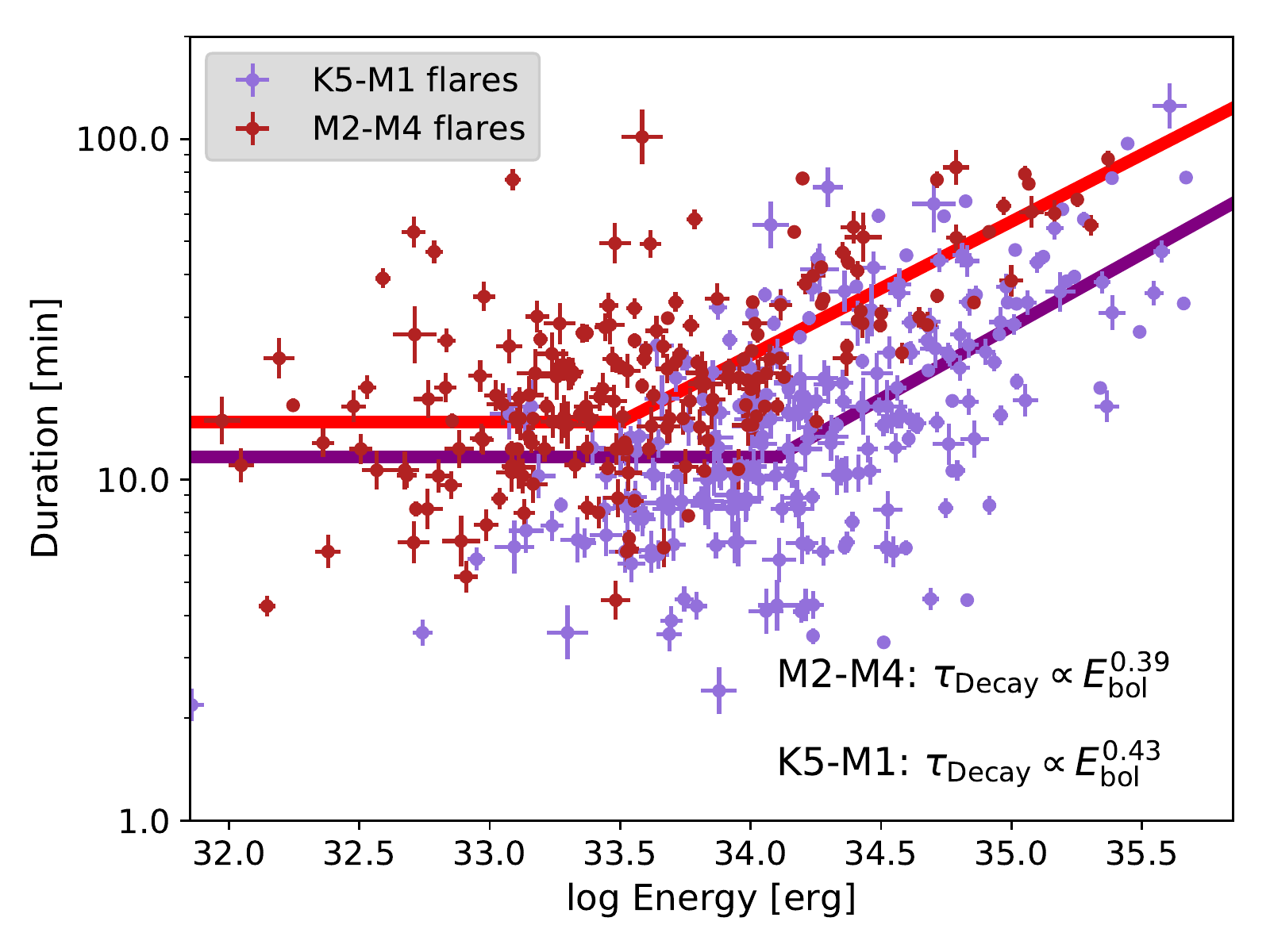}
	}
	\caption{Top panel: Flare energy and duration (i.e. decay timescale) of all flares discovered by the automated pipeline in Section \ref{evr_auto_elfs}. Errors in energy and duration are computed as the inverse significance of detection. A broken power law is fit, with photometric scatter dominating below 10$^{34}$ erg. At energies above 10$^{34}$ erg, the best-fit power law coefficient of 0.34 is consistent with that predicted by magnetic re-connection, 1/3. Bottom panel: Same as top panel, but with separate power laws fit to early M and late K flares. We observe a gradient in the flare duration at a given energy as a function of spectral type. Both power law coefficients are slightly larger than 1/3, although it is unclear if this is due entirely to the large scatter in the data or implies emission mechanisms beyond magnetic re-connection.}
	\label{fig:mag_reconnect}
\end{figure}

\begin{figure*}
	\centering
    \subfigure
	{
		\includegraphics[trim= 0 0 0 0, clip, width=2.25in]{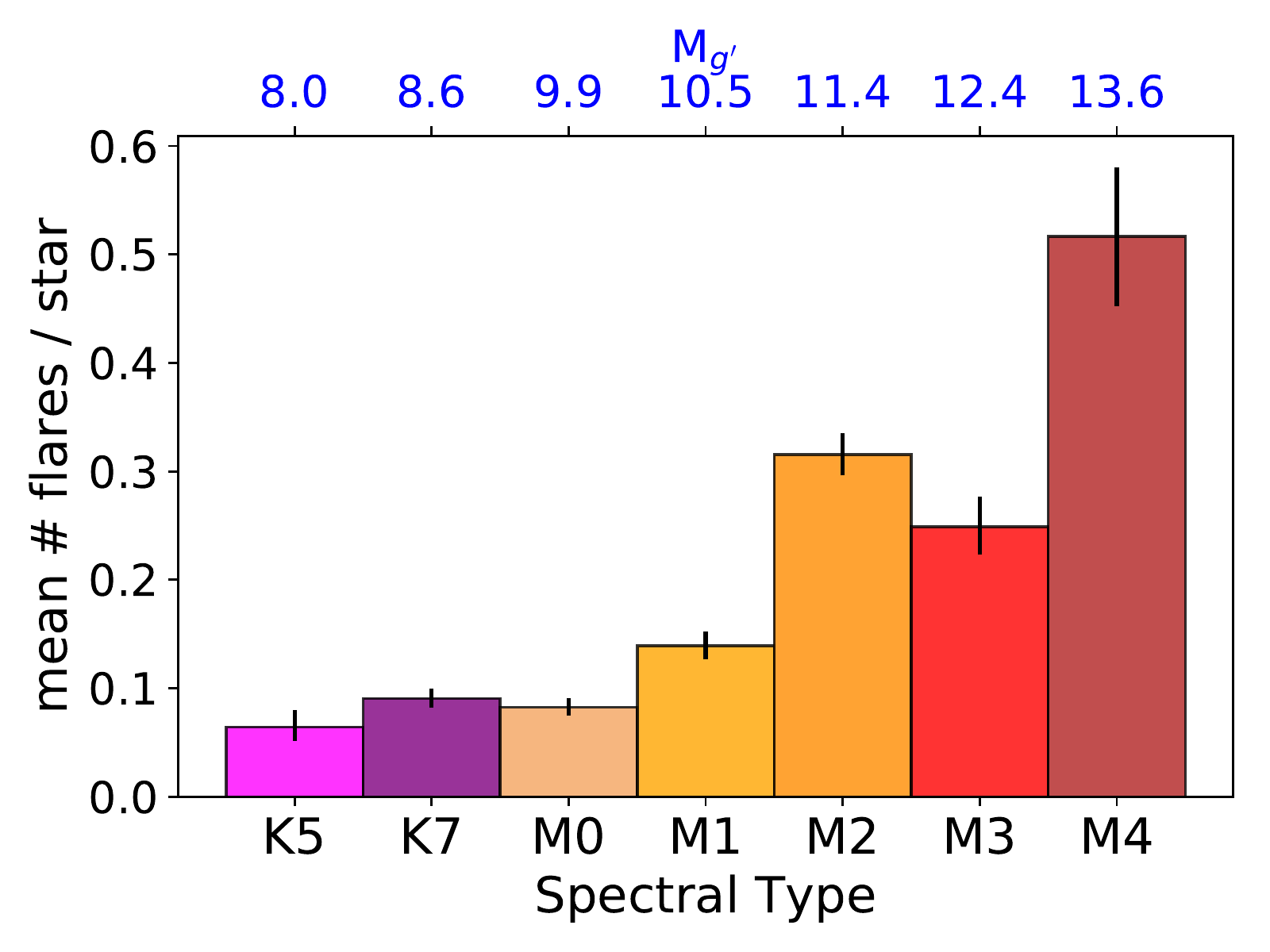}
		\label{fig:indiv_flares_vs_spt_threepanel}
	}
	\subfigure
	{
		\includegraphics[trim= 0 0 0 0, clip, width=2.25in]{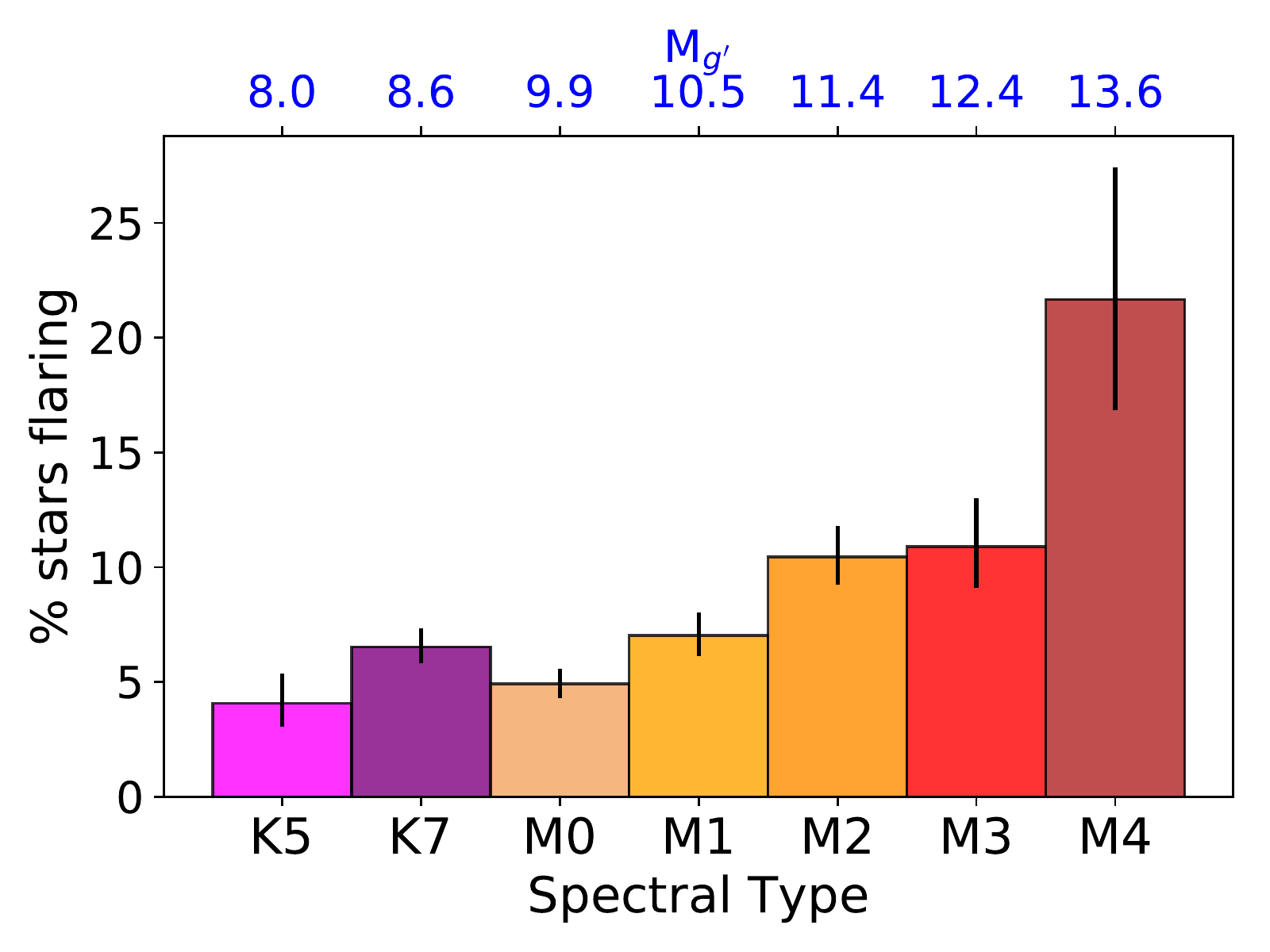}
		\label{fig:flare_stars_vs_spt_threepanel}
	}
	\subfigure
	{
		\includegraphics[trim= 0 0 0 0, clip, width=2.25in]{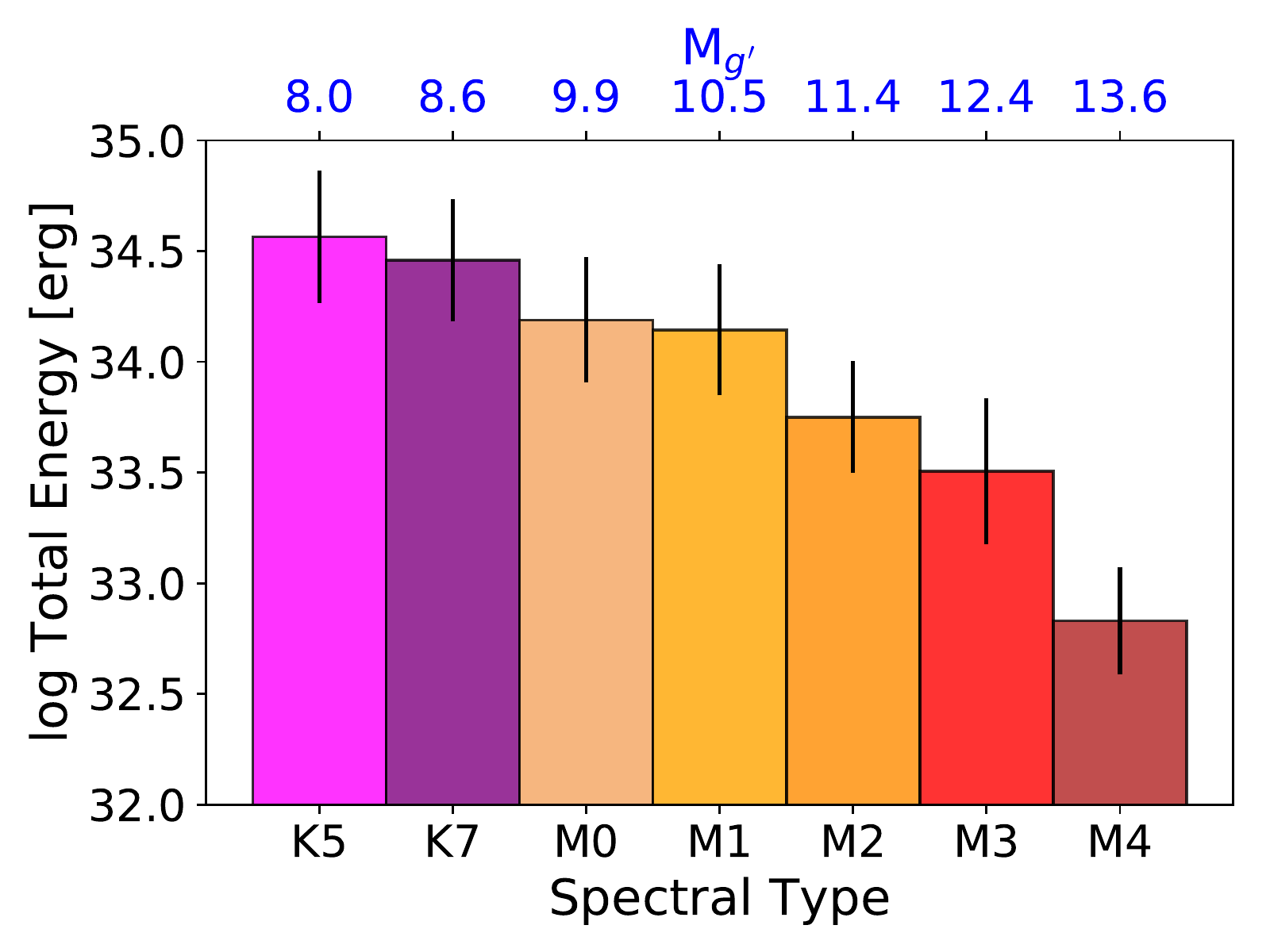}
		\label{fig:flare_energy_vs_spt}
	}
	\caption{Flaring as a function of spectral type. Left panel: the average number of individual flares observed per star as a function of spectral type. Error bars are 1$\sigma$ binomial confidence intervals. Middle panel: the fraction of flare stars observed as a function of spectral type. Error bars are 1$\sigma$ binomial confidence intervals.  We note a rise in the average number of flares and the fraction of flare stars towards the M4 fully-convective boundary. Right panel: the flare energy as a function of spectral type. Error bars are the standard deviation in energy.}
	\label{fig:spt_trends_flares}
\end{figure*}

\begin{figure}
	\centering
	{
		\includegraphics[trim= 10 0 0 12, clip, width=3.4in]{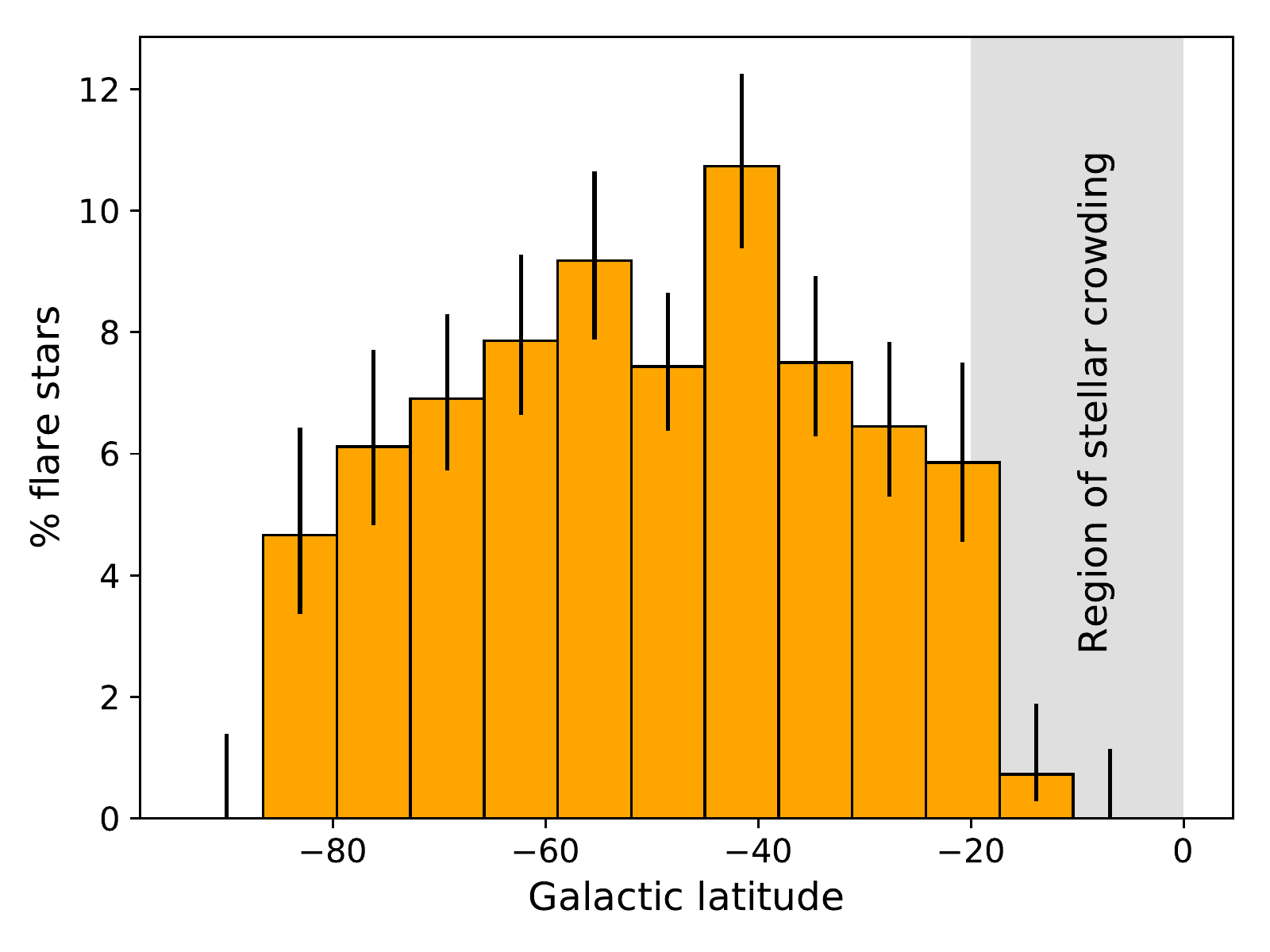}
	}
	\caption{The percentage of flare stars in our sample of cool stars is displayed as a function of galactic latitude. Error bars are 1$\sigma$ binomial confidence intervals. We note an apparent decrease in the flare rate at high galactic latitudes. This may be due to the decreased activity of old stars above the galactic plane; it may also be a result of sampling the decreasing density of both flaring and non-flaring M-dwarfs at high latitudes.}
	\label{fig:glat_flare_stars}
\end{figure}

\section{Evryscope flare discoveries}\label{evr_flare_distrib}
We detect 575 high-energy flare events from 284 flare stars in TESS sectors 1-6. Such a large sample of high-energy flares from cool stars probes both the dependence of superflaring on other astrophysical parameters and the potential habitability of planets orbiting cool star stars: we present $\sim2\times$ the previous-largest sample of high-cadence 10$^{34}$ erg flares from nearby cool stars (e.g. \citet{gunter2019}).

We detect at least an order of magnitude more large flares than other ground-based flare surveys due to the high-cadence and multi-year coverage of the entire accessible sky. Precision Evryscope light curves of flare stars later than M4 are only possible for the brightest sources across the sky, although this does not rule out the detection of multi-magnitude flare events from late M and L dwarfs in future work using a separate pipeline. In comparison, the large flare yield of ASAS-SN displayed in Figure 1 of \citet{Schmidt_2018_ASASSN_flares} increases significantly at later types.

\subsection{Flare stars, spectral type, and stellar age}\label{evr_indiv_flares_spt}
We explore how superflare rates correlate with drivers of stellar surface magnetic activity. 

\subsubsection{Superflare energy and duration}\label{evr_mag_reconnect}
Because flares are thought to result from magnetic re-connection, we begin by attempting to confirm that our very large superflare events distribute their energy release according to the predictions of magnetic re-connection models. \citet{Namekata2017} describe how flares generated by magnetic re-connection follow the scaling relation $\tau_\mathrm{decay}\propto E_\mathrm{bol}^{1/3}$ between flare energy $E_\mathrm{bol}$ and flare duration $\tau_\mathrm{decay}$ (i.e. the approximate decay time). Our distribution of flare energy versus duration shown in the top panel of Figure \ref{fig:mag_reconnect} follows a broken power law that is flat at energies below 10$^{34}$ erg and best fit by $\tau_\mathrm{decay}\propto E_\mathrm{bol}^{0.34}$ above this energy. The flat power law at lower energies is due to the flare decay tail falling below the photometric noise level and biasing the measured duration. However, when we split up our flares into late K and early M bins and re-compute the power laws separately as shown in the bottom panel of Figure \ref{fig:mag_reconnect}, we observe coefficients of $\sim$0.4, slightly larger than those expected from re-connection. We estimate our broken power law knee in flare energy to be approximately 10$^{33.5}$ for late K flares and 10$^{34.1}$ for early M flares.

In fact, our coefficients for the late K and early M bins are within the errors of the G-dwarf superflare coefficient measurement of 0.38$\pm$0.06 discussed in \citet{Namekata2017}. \citet{Namekata2017} also considers a number of additions to magnetic re-connection that may steepen the power law. Because the scatter in the data is large, it is unclear whether the larger coefficients we find for the separate populations imply emission mechanisms beyond magnetic re-connection.

We note that we exclude durations measured by hand in the manual pipeline of Section \ref{evr_flare_search} due to bias in the measured flare start and stop times (increased to ensure the flare fell between the selected times). We conclude that our superflares are broadly consistent with being generated by the re-connection process, but may be affected by additional mechanisms, as in \citet{Namekata2017}.

\begin{figure*}
	\centering
	{
		\includegraphics[trim= 1 1 1 1,clip, width=6.9in]{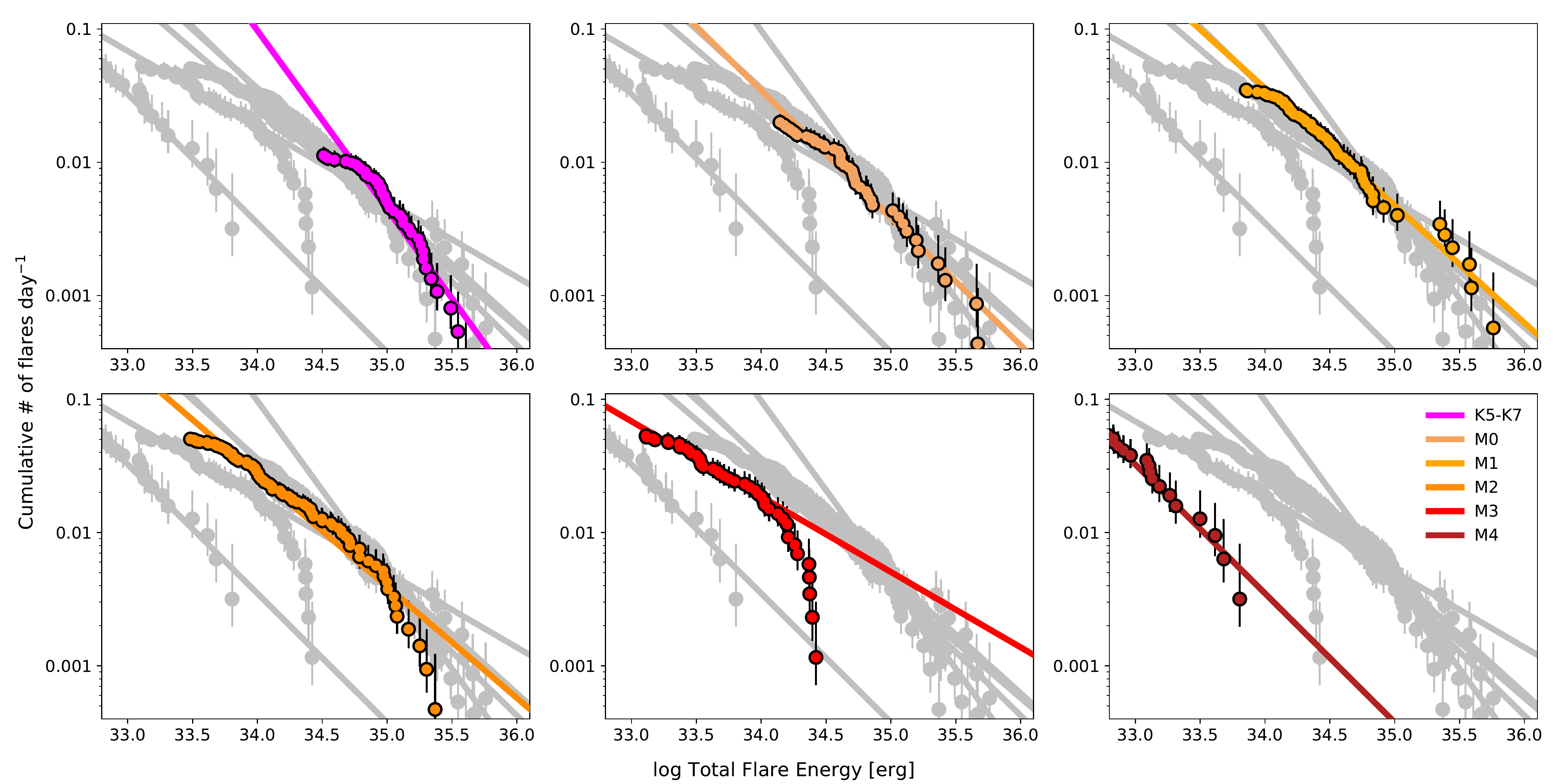}
	}
	\caption{We construct averaged cumulative FFDs for each spectral type classification. We bin all flares observed and the total observing time by the estimated spectral types. As a result, these relations do not hold for inactive stars. Errors in the number of flares d$^{-1}$ are given by 1$\sigma$ binomial confidence intervals. The curve at the lower-energy end of each FFD is an artifact of sometimes failing to observe the smallest flares.We remove all flares with an ED$<$10$^{2.44}$ from the fit, below which the lost flares dominate. Because this incompleteness limit is higher for later types, this curve remains visible at the leftmost end of each panel.}
	\label{fig:average_ffd_per_spt}
\end{figure*}

\begin{table*}
\caption{Flare wait-times and FFD fit parameters for average K5-M4 flare stars}
\begin{tabular}{p{2.4cm} p{2.4cm} p{2.4cm} p{2.4cm} p{2.4cm} p{2.4cm}}
\hline
 & & & & & \\
SpT & $\alpha_\mathrm{energy}$ & $\beta_\mathrm{energy}$ & Max energy \newline seen in 10 d & Max energy \newline seen in 28 d & Waiting-time \newline for 10$^{33}$ erg flare\\
 & &  & [log erg]  & [log erg] & [d] \\
\hline
 & & & & & \\
Active K5 & -1.34 & 44.55 & 34.0 & 34.3 & 0.5\\
Active K7 & -1.34 & 44.55 & 34.0 & 34.3 & 0.5\\
Active M0 & -0.96 & 31.05 & 33.5 & 34.0 & 3.2\\
Active M1 & -0.88 & 28.5 & 33.5 & 34.0 & 3.7\\
Active M2 & -0.84 & 26.82 & 33.3 & 33.9 & 5.4\\
Active M3 & -1.25 & 40.02 & 32.9 & 33.3 & 12.0\\
Active M4 & -0.97 & 30.45 & 32.5 & 33.0 & 30.8\\
 &  &  &  &  & \\
\hline
\end{tabular}
\label{table:energyFFD_tab}
{\newline\newline \textbf{Notes.} Fit parameters to the ``averaged" FFD for K5-M4 flare stars, shown in Figure \ref{fig:average_ffd_per_spt}. $\alpha$ and $\beta$ are given by the power law of the form $\log{\nu}=\alpha\log{E} + \beta$ as described in Section \ref{evry_flare}, where $\nu$ is the number of flares observed per day at an energy of at least $E_mathrm{bol}$. We estimate the largest flare expected from a typical active star of each spectral type during 10 and 28 days of continuous observing, respectively. We also estimate the waiting-time between successive flares of at least 10$^{33}$ erg.}
\end{table*}

\subsubsection{Flare Frequency vs. Spectral Type and Galactic Latitude}
Superflare energy and occurrence will impact the atmospheres of temperate planets differently depending on the host star's spectral type. We use M$_{g'}$ to estimate the spectral type of each flare star. Due to the faintness of stars later than M4 in the blue, we do not include later types in this analysis.

Both the average number of flares per star and the fraction of searched stars that flare increase from K7 toward M4. This may be a result of approaching the fully-convective boundary. We define the average number of flares per star per spectral type as the number of flares observed from all stars of a given spectral type divided by the total number of stars of that spectral type in our flare search. Error bars are given by 1$\sigma$ binomial confidence intervals for each spectral type in the two panels to the left in Figure \ref{fig:spt_trends_flares}.

Remarkably, the fraction of cool flaring stars per spectral type is identical to the fraction of flaring M-dwarfs found at lower flare energies in \citet{gunter2019}, indicating that superflares from late-type stars follow a similar increase in flare activity as small flares. The fraction of flaring M-dwarfs per spectral type is also comparable to that found by \citet{Yang2017} in \textit{Kepler} light curves. The fraction of active stars for each spectral type measured in \citet{West2008,west2015} and \citet{Schmidt_2018_ASASSN_flares} are 2-10X as high as those we measure here. This is likely a result of choosing to measure activity using a sample of infrequent superflares rather than elevated H$\alpha$ emission in spectra.

We also check if the occurrence of large flares depends upon galactic latitude. Stars in the disk are generally younger and therefore more active than stars at higher latitudes \citet{West2008}. In Figure \ref{fig:glat_flare_stars}, we do observe an apparent decrease in flare stars at high latitudes. This may be due in part to target selection, as there are fewer cool stars at high latitudes than low latitudes.

Flare surveys in \textit{Kepler} and K2 data (\citet{Borucki2010,Howell2014}) also find increases in flare rate and fraction of stars flaring for K5 and later spectral types, and decreased flaring with greater age  across spectral type  \citep{Candelaresi2014,Davenport2016_catalog,Doorsselaere2017,Yang2017,Davenport2019,Ilin2019}. However, \textit{Kepler} and K2 only observed several hundred active M-dwarfs (e.g. \citet{Davenport2016_catalog,Stelzer2016}). Evryscope and TESS observations of orders-of-magnitude more M-dwarfs will provide comprehensive flare monitoring in the M-dwarf regime \citep{gunter2019}.

Several caveats are in order: Figure \ref{fig:spt_trends_flares} gives the occurrence of the largest flares; surveys observing smaller flares may therefore observe higher rates of flaring. Next, the increased flaring of M4 dwarfs involves small-number statistics. Although larger than for other spectral types, M4 errors are still $<$20\%. Last, we do not perform flare injection and recovery in this sample, so Evryscope systematics in the light curves could alter the true number of stars from which we would have been able to see flares. Because 10\% of Evryscope light curves experience source contamination from stellar crowding outside the galactic plane, we conclude this is not a dominant source of error.

\subsubsection{Mean Flare Energy vs. Spectral Type}
Next, we find that the mean flare energy decreases as a function of spectral type, as shown in the right panel of Figure \ref{fig:spt_trends_flares}. Error bars are the 1$\sigma$ spread in energy. As \citet{gunter2019} and \citet{Davenport2019} note, the lower luminosity of the later types means the same ED results in less bolometric energy. In \textit{Kepler}, the maximum flare energy as a function of spectral type shows a similar decline toward later types \citep{Davenport2016_catalog}.

We compute the hemispherical starspot coverage necessary to generate flares at the mean flare energies we observe for each spectral type in Figure \ref{fig:spt_trends_flares} as described later in Section \ref{spots}. We assume a stellar magnetic field of 1 kG and compute starspot coverage as described in Section \ref{spots}. We find that spots corresponding to the observed mean flare energies cover 1-2\% of the stellar hemisphere across all K5-M4 spectral types.

Although the mean flare energy of late K and early M stars in our sample is high, future work is needed to determine if the increased orbital distance to the HZ will protect the atmospheres of Earth-like planets around these stars.

\subsubsection{Superflare Rate vs. Spectral Type}\label{average_FFD_section}
\begin{figure}
	\centering
	{
		\includegraphics[trim= 1 9 0 12, clip, width=3.4in]{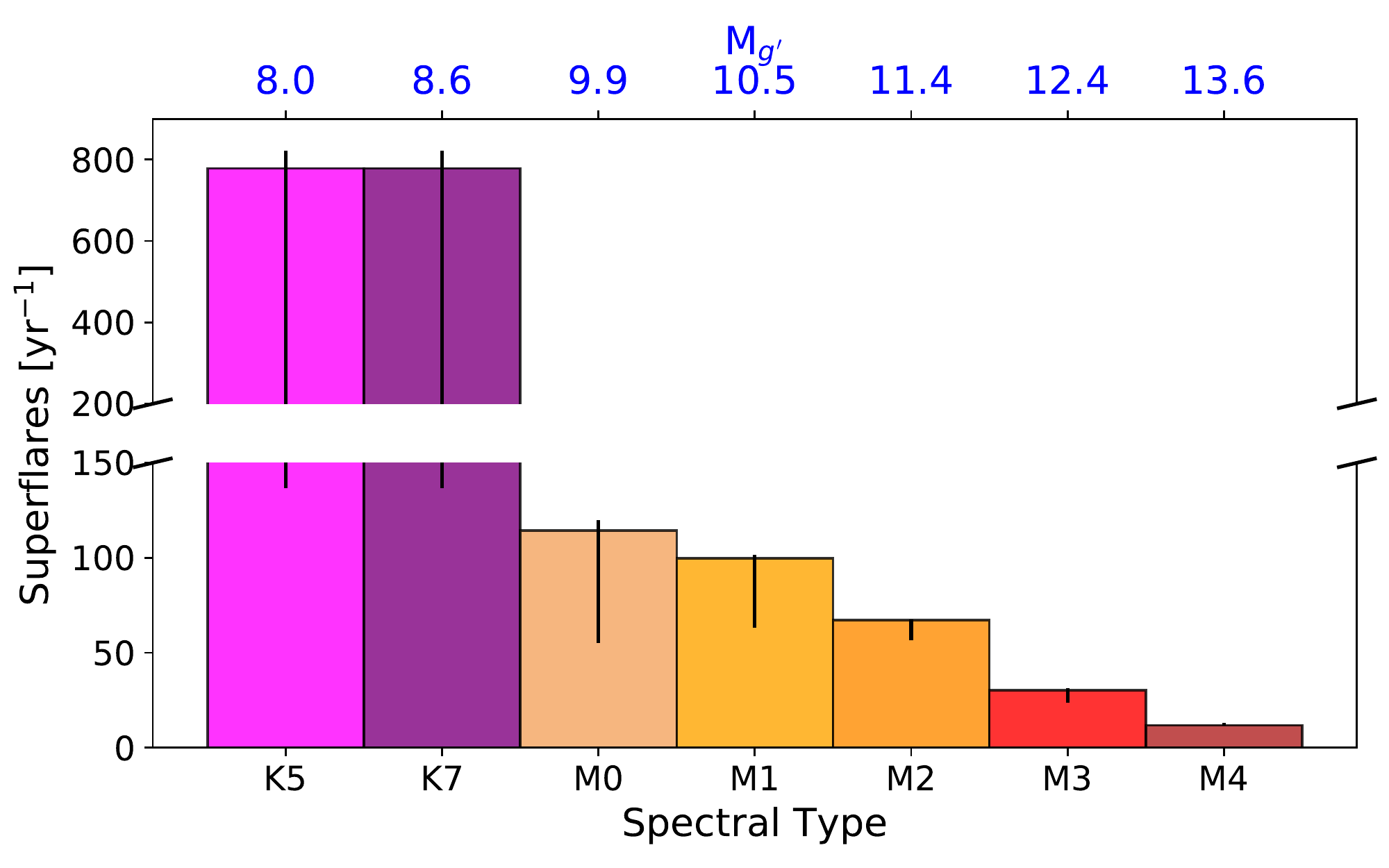}
	}
	\caption{The annual superflare rate of a typical active flare star as a function of estimated spectral type. We extrapolate the superflare rate from each averaged cumulative FFD for each spectral type displayed in Figure \ref{fig:average_ffd_per_spt}. As a result, this distribution does not hold for inactive stars. Due to the low numbers of K-dwarf flares, we bin all K5-K7 flares and display the averaged result in both the K5 and K7 bins for consistency with other plots in this work. Error bars on superflare rates are calculated with 1000 posterior draws to each FFD.}
	\label{fig:annual_superflares}
\end{figure}

To investigate superflare frequency, we construct cumulative FFDs for an ``average" flaring star of each spectral type. Binning all flare observations by spectral type, we find similar power-law slopes $\alpha$ for early and mid-M stars, but higher y-intercepts $\beta$ and therefore occurrence of flares at a given energy for the earlier types. The FFDs are displayed in Figure \ref{fig:average_ffd_per_spt}. We estimate the annual rate of 10$^{33}$ erg superflares in Figure \ref{fig:annual_superflares}. We record the fitting functions of each FFD and the expected waiting-times for a superflare to occur in Table \ref{table:energyFFD_tab}.

Although these rates are high, they are constructed from active stars of each spectral type and do not hold for inactive stars. \citet{Loyd2018} finds inactive stars to be $10\times$ less active in the FUV-130 bandpass. Should similar relationships hold for white-light superflares, the impacts on planet atmospheres would be greatly reduced for inactive stars.

\subsubsection{High-amplitude Flare Occurrence vs Spectral Type}\label{high_ampl_indiv_flares_spt}
\begin{figure*}
	\centering
	{
		\includegraphics[trim= 1 1 1 1,clip, width=6.9in]{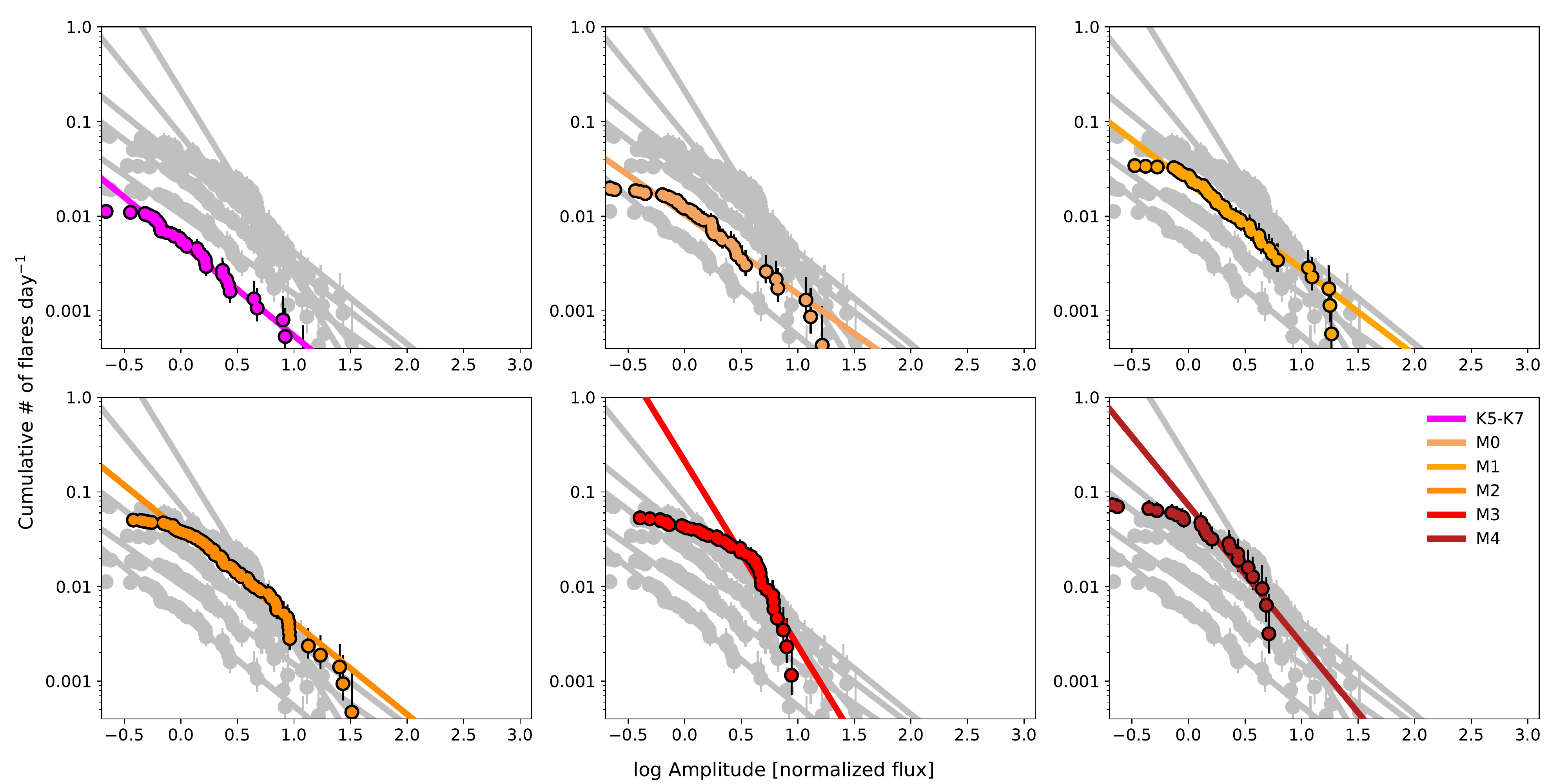}
	}
	\caption{We construct averaged cumulative FFDs from flare amplitudes instead of flare energies. We bin all flare amplitudes observed and the total observing time by the estimated spectral types. As a result, these relations do not hold for inactive stars. Errors in the number of flares d$^{-1}$ are given by 1$\sigma$ binomial confidence intervals. The curve at the lower end of each FFD is an artifact of sometimes failing to observe the smallest flares. We remove all flares with an ED$<10^{2.44}$ from the fit, below which the lost flares dominate. Because this incompleteness limit is higher for later types, this curve remains visible at the leftmost end of each panel (strongest for M3). We further manually adjust the fit to include only the linear-in-log-log region of the M3 and M4 amplitude power laws to avoid bias at the lower end.}
	\label{fig:average_peakffd_per_spt}
\end{figure*}

\begin{table*}
\caption{Flare wait times and flare amplitudes-occurrence FFD fit parameters for average K5-M4 flare stars}
\begin{tabular}{p{1.8cm} p{1.4cm} p{1.4cm} p{1.9cm} p{1.9cm} p{1.9cm} p{1.9cm} p{2.4cm}}
\hline
 & & & & & \\
SpT & $\alpha_\mathrm{ampl}$ & $\beta_\mathrm{ampl}$ & Max ampl.\newline seen in 10 d & Max contrast seen in 10 d \newline & Max ampl.\newline seen in 28 d  & Max contrast\newline seen in 28 d & Waiting-time\newline for 3-mag flare\\
 & & & [$\Delta$F/F] & [$\Delta$M$_{\textit{g}\textsuperscript{$\prime$}}$] & [$\Delta$F/F] & [$\Delta$M$_{\textit{g}\textsuperscript{$\prime$}}$] & [yr]\\
\hline
 &  &  &  &  &  &  & \\
Active K5 & -0.44 & -2.34 & 0.0 & 0.0 & 0.0 & 0.0 & N/A\\
Active K7 & -1.04 & -2.26 & 0.1 & 0.1 & 0.2 & 0.2 & 8.2\\
Active M0 & -0.84 & -1.98 & 0.1 & 0.1 & 0.2 & 0.2 & 2.5\\
Active M1 & -0.91 & -1.65 & 0.2 & 0.2 & 0.6 & 0.5 & 1.4\\
Active M2 & -0.97 & -1.41 & 0.4 & 0.3 & 1.1 & 0.8 & 1.0\\
Active M3 & -2.19 & -0.69 & 1.4 & 0.9 & 2.2 & 1.3 & 4.9\\
Active M4 & -1.46 & -1.14 & 0.8 & 0.6 & 1.6 & 1.0 & 2.0\\
 &  &  &  &  &  &  & \\
\hline
\end{tabular}
\label{table:amplFFD_tab}
{\newline\newline \textbf{Notes.} Fit parameters to the ``averaged" flare amplitudes ``FFD" for K5-M4 flare stars, shown in Figure \ref{fig:average_peakffd_per_spt}. $\alpha$ and $\beta$ are given by the power law of the form $\log{\nu}=\alpha\log{A} + \beta$ following the discussion in Section \ref{high_ampl_indiv_flares_spt}, with $\nu$ being the number of flares observed per day at an amplitude with a fractional flux peak of at least $A$. We estimate the largest flare amplitude expected from a typical active star of each spectral type during 10 and 28 days of continuous observing, respectively. Each amplitude is given in units of both fractional flux and $\textit{g}^{\prime}$ magnitudes. We also estimate the waiting-time between successive flares of at least 3 \textit{g}$^{\prime}$ magnitudes.}
\end{table*}

Sky surveys performing rapid transient discovery and follow-up must be able to characterize the degree to which M-dwarf flare stars contaminate desired triggers from extra-galactic sources of rapid brightening events (e.g. \citet{Ho2018,Andreoni2019,Roestel2019}). 

We construct cumulative FFDs for flare amplitudes rather than energies in order to predict how often an average flare star of a given spectral type will emit a flare of a given amplitude. We fit parameters $\alpha$ and $\beta$ to the power law $\log{\nu}=\alpha\log{A} + \beta$ following the discussion in Section \ref{high_ampl_indiv_flares_spt}, with $\nu$ being the number of flares observed per day at an amplitude with a fractional flux peak of at least $A$. Recorded in Table \ref{table:amplFFD_tab} and displayed in Figure \ref{fig:average_peakffd_per_spt}, the resulting amplitude-FFDs may be used to predict how often a flare of a given amplitude will occur, as well as the largest flare expected within a certain observing baseline. For example, a survey observing an M2e star for 10 continuous days would observe a flare with a stellar peak fractional flux of least 0.4.

\begin{figure}
	\centering
	{
		\includegraphics[trim= 1 0 0 12, clip, width=3.4in]{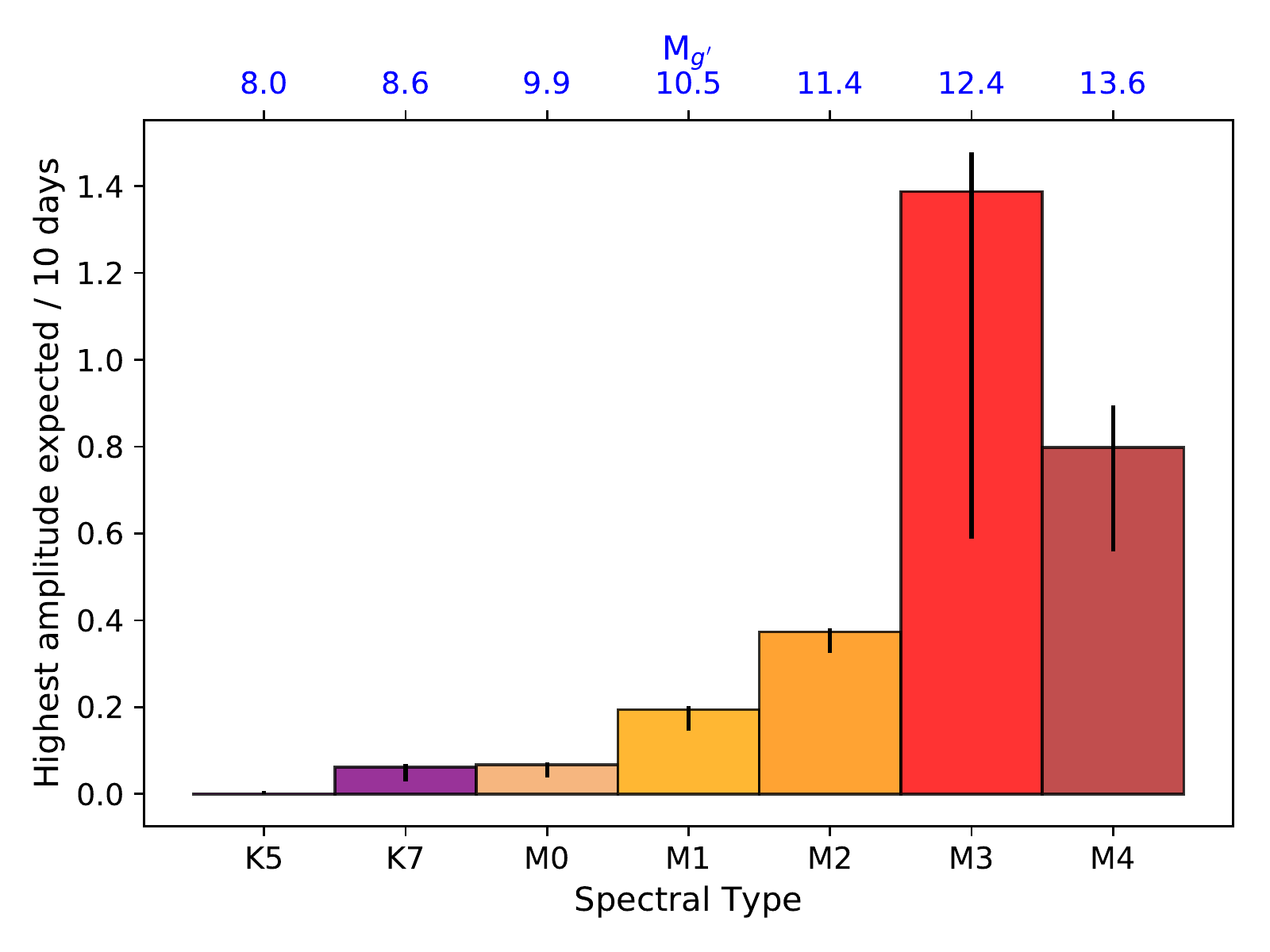}
	}
	\caption{We estimate the largest flare amplitude expected from a typical active star of each spectral type during 10 days of continuous observing. These estimates are obtained by extending the amplitude FFDs in Figure \ref{fig:average_peakffd_per_spt} and Table \ref{table:amplFFD_tab} to the typical flare amplitude per 10 days of observing. Flare amplitudes are displayed as peak increases in fractional flux. We find expected amplitudes increase for less luminous spectral types. The large uncertainty in M3 is due to the knee in the power law in Figure \ref{fig:average_peakffd_per_spt}.}
	\label{fig:superflare_ampl_in_10d_vs_spt}
\end{figure}

The best-fit parameters for these amplitude-FFDs for each spectral type are given in Table \ref{table:amplFFD_tab}. We find the largest flare amplitude expected from a typical active star of each spectral type increases as the quiescent luminosity of the star decreases as shown in Figure \ref{fig:superflare_ampl_in_10d_vs_spt}. We also find the waiting-time between successive flares of at least 3 \textit{g}$^{\prime}$ magnitudes decreases from nearly a decade for late K-dwarfs to only two years for M4 dwarfs.

\subsubsection{Starspot coverage and superflares}\label{spots}
41 of our flare events exceed 10$^{35}$ erg. If the energy released by extreme flares is stored in surface magnetic fields, then the area of the smallest spot that could have produced such a flare is given by $E_\mathrm{flare}=\frac {B^2}{8\pi} A_\mathrm{spot}^{3/2}$ \citep{Shibata2013,Notsu2019}. E$_\mathrm{flare}$ is the bolometric flare energy, B is the surface magnetic field strength, and A$_\mathrm{spot}$ is the smallest spot group area expected to generate E$_\mathrm{flare}$. We note that this model is a very simplified assumption and true spot sizes could be at least an order of magnitude larger. We estimate the starspot coverage by dividing A$_\mathrm{spot}$ by the projected area of the approximate stellar hemisphere A$_\mathrm{star}$. To calculate stellar area, we estimate the stellar mass from M$_{g'}$ using \citet{Kraus2007} and then estimate the stellar radius using the mass-radius relationship provided by \citet{Mann2015}. We note that the spot group area scaling law was discovered for solar-type stars and extrapolation into the cool star regime may introduce further error.

\begin{figure}[t]
	\centering
	{
		\includegraphics[trim= 1 0 0 12, clip, width=3.4in]{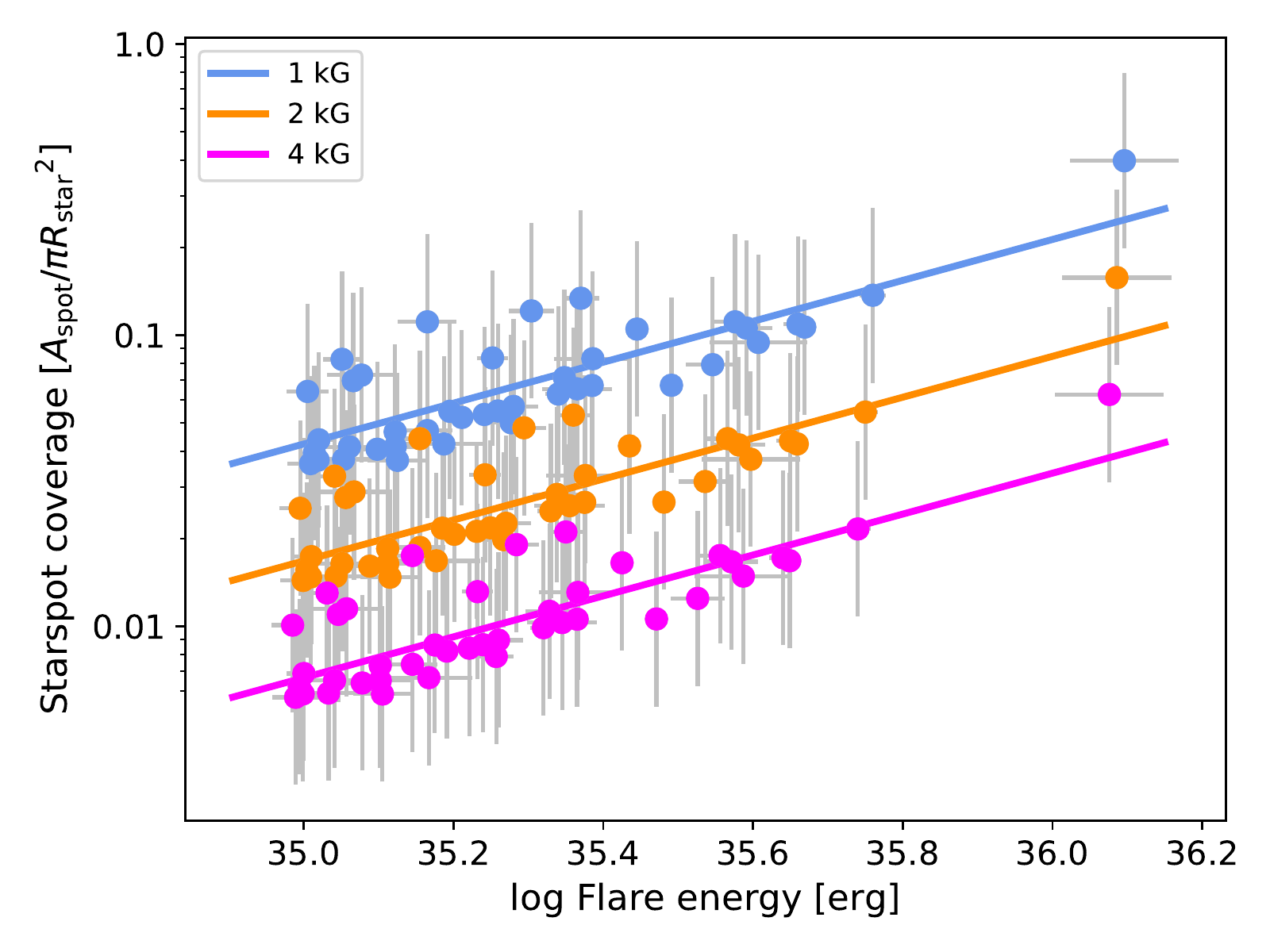}
	}
	\caption{Estimated starspot coverage required to generate the largest superflares we observed as a function of stellar magnetic field strength and flare energy. We compute the starspot coverage as the spot group area divided by the projected area of the stellar hemisphere. The minimum spot group area required to generate each superflare is computed from the flare energy using scaling relations from \citet{Shibata2013,Notsu2019}; \textbf{true spot coverage could be at least an order of magnitude larger.} We fit a power law of the form $\log{f_\mathrm{coverage}}=a\log{E} + b$ to the spot coverage $f_\mathrm{coverage}$ and flare energy $E$ for representative cool star field strengths. Fit coefficients are given in Table \ref{table:Bfield_tab}. Error in energy is computed as the inverse significance of detection; 100\% error in spot coverage is assumed due to the approximate nature of the spot group area scaling law.}
	\label{fig:superflare_spot_coverage}
\end{figure}

Early to mid M-dwarf surface magnetic fields are often 1-4 kG in strength \citep{Shulyak2017}. We therefore estimate the approximate starspot coverage as a function of flare energy for 1, 2, and 4 kG fields, shown in Figure \ref{fig:superflare_spot_coverage}. As the field strength increases, the necessary spot coverage to generate a given superflare decreases \citep{Notsu2019}. We compute three separate scaling relationships between flare energy and starspot coverage of K5-M4 stars for 1, 2, and 4 kG fields assuming a power law of the form $\log{f_\mathrm{coverage}}=a\log{E} + b$, where $f_\mathrm{coverage}$ is the spot coverage and $E$ is the flare energy. The fits are also shown in Figure \ref{fig:superflare_spot_coverage}.

We attempt to constrain the largest flare a cool star may emit by assuming 100\% hemispherical spot coverage and solving for the flare energy. The hypothesized maximum-allowed flare energies are displayed in Table \ref{table:Bfield_tab} along with an estimate of the waiting-time between successive flares at these energies obtained from the K5-M4 FFDs in Section \ref{average_FFD_section}. We caution readers that these upper limits are dependent on large uncertainties in the flare energy-spot model and in the FFDs. Cool star flare energies associated with 100\% spot coverage are comparable to those estimated by \citet{Notsu2019} for Solar-type stars.

\begin{table*}
\caption{Starspot coverage of average K5-M4 flare stars}
\begin{tabular}{p{1.1cm} p{0.9cm} p{0.9cm} p{1.8cm} p{1.6cm} p{1.6cm} p{1.6cm} p{1.6cm} p{1.6cm} p{1.6cm}}
\hline
 & & & & & \\
Stellar \newline B-field \newline strength & $a_\mathrm{spot}$ & $b_\mathrm{spot}$ & Max allowed \newline flare energy, \newline $E_\mathrm{max}$ & K5-K7 \newline Wait-time \newline for $E_\mathrm{max}$ & M0 \newline Wait-time \newline for $E_\mathrm{max}$ & M1 \newline Wait-time \newline for $E_\mathrm{max}$ & M2 \newline Wait-time \newline for $E_\mathrm{max}$ & M3 \newline Wait-time \newline for $E_\mathrm{max}$ & M4 \newline Wait-time \newline for $E_\mathrm{max}$ \\
 & & & & & \\
~[kG] &  &  & [log erg] & [kyr] & [kyr] & [kyr] & [kyr] & [kyr] & [kyr] \\
\hline
 & & & & & \\
1 kG & 0.70 & -25.96 & 37.0 & 0.3 & 0.07 & 0.03 & 0.05 & 4 & 0.7 \\
2 kG & 0.70 & -26.36 & 37.5 & 1.5 & 0.3 & 0.09 & 0.1 & 20 & 2.4\\
4 kG & 0.70 & -26.76 & 38.1 & 8.7 & 0.9 & 0.3 & 0.4 & 110 & 8.8\\
 &  &  &  &  & \\
\hline
\end{tabular}
\label{table:Bfield_tab}
{\newline\newline \textbf{Notes.} Fit coefficients for the power law of the form $\log{f_\mathrm{coverage}}=a\log{E} + b$ describing the scaling relationship between hemispherical spot coverage of cool stars $f_\mathrm{coverage}$ and superflare energy $E$. We perform separate fits at representative cool star magnetic field strengths. We also hypothesize for each field strength the maximum allowed flare energy $E_\mathrm{max}$ assuming 100\% spot coverage. We urge caution in applying these maximum-allowed flare energies, because real spots do not necessarily release all of their energy in a single flare event. As a result, the flare energy and spot size scaling used to compute these values introduces at least order-of-magnitude-level uncertainties. Using the FFDs computed in Table \ref{table:energyFFD_tab}, we estimate the waiting-time between successive flares of energy $E_\mathrm{max}$ for an active star in each spectral type.}
\end{table*}

\subsection{Comparing Evryscope and TESS flares}\label{evr_ampl_energies}
Evryscope flare monitoring of TESS flare stars complements flare studies done in the TESS light curves themselves. While TESS has the high photometric precision necessary to observe the most frequent low-to-moderate energy flares, long-term Evryscope monitoring captures the largest and rarest flares the star is capable of releasing. 

\begin{figure*}
	\centering
    \subfigure
	{
		\includegraphics[trim= 10 0 10 30, clip, width=3.4in]{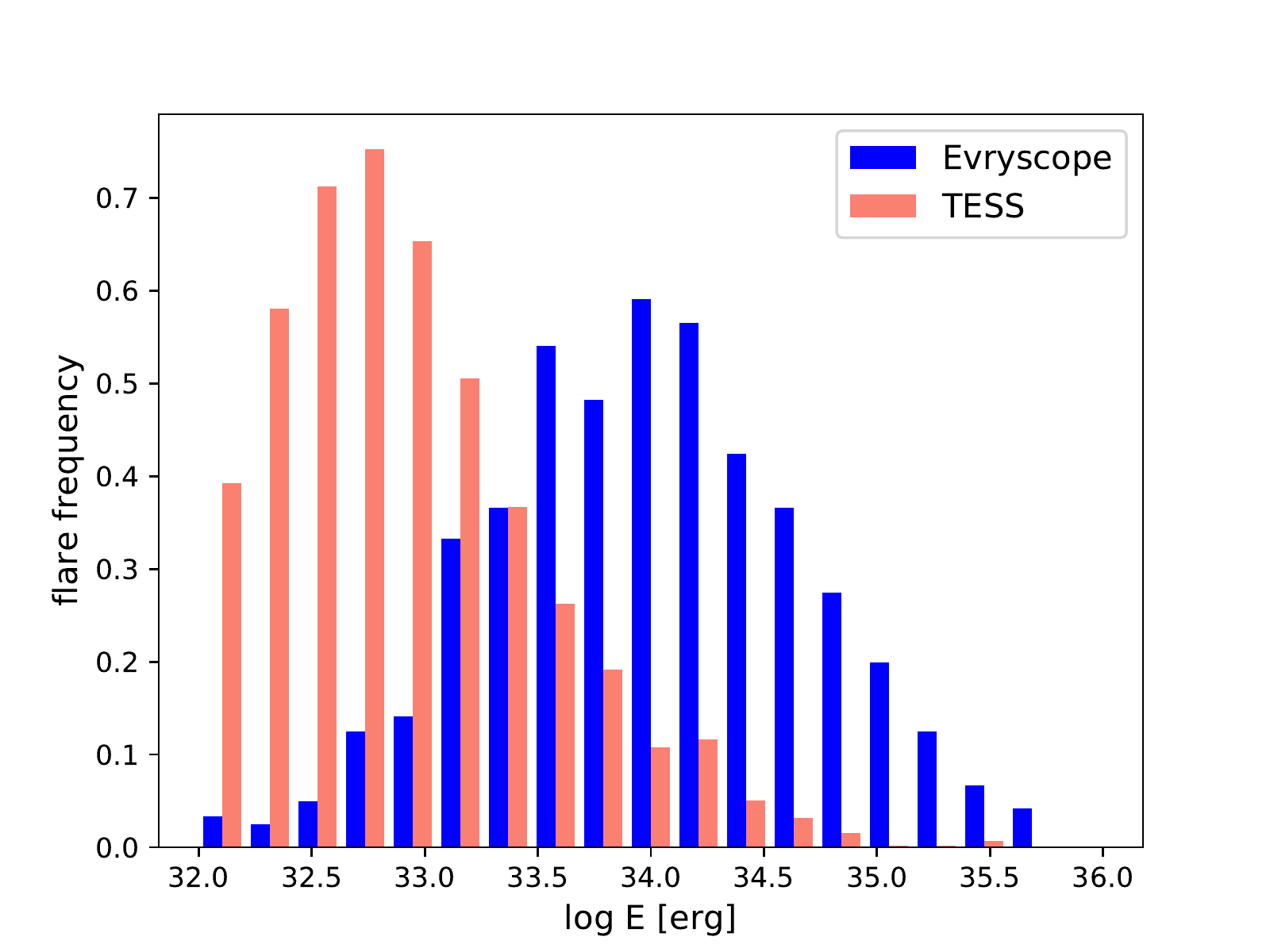}
		\label{fig:Evryscope_vs_TESS_a}
	}
	\subfigure
	{
		\includegraphics[trim= 0 0 -20 0, clip, width=3.4in]{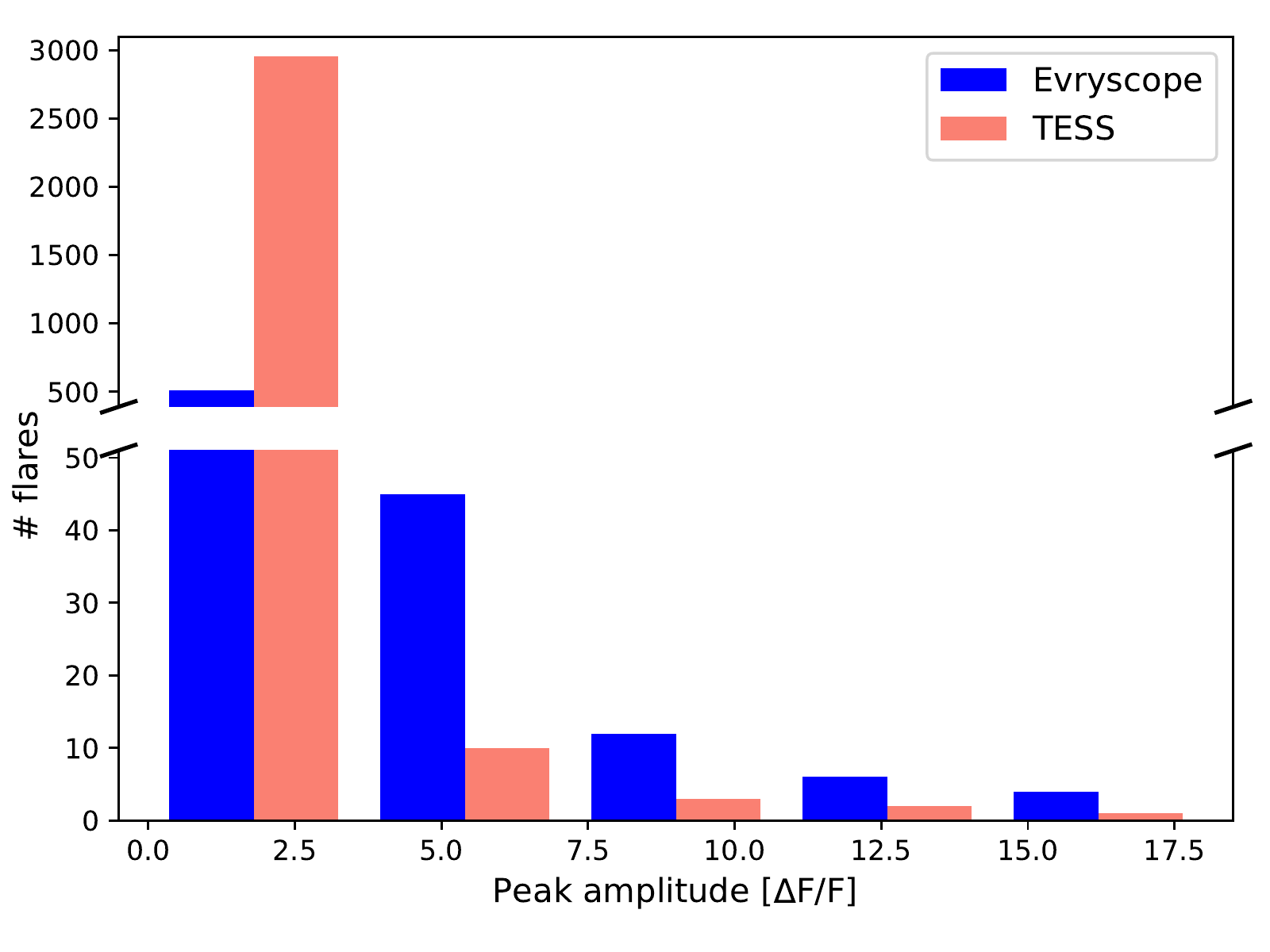}
		\label{fig:Evryscope_vs_TESS_b}
	}
	\caption{Left panel: Normalized distributions of recovered bolometric flare energies from Evryscope and TESS. Right panel: Histogram of flare amplitudes from Evryscope and TESS. TESS flares in both panels are from light curves of K5 and later stars in \citet{gunter2019}. Although TESS observes an order of magnitude more flares, Evryscope captures the largest-amplitude and highest-energy flare events.}
	\label{fig:Evryscope_vs_TESS}
\end{figure*}

Flares observed by Evryscope are approximately an order of magnitude more energetic than those found in the TESS light curves themselves due to the longer observing baseline and lower photometric precision of Evryscope compared to TESS, as displayed in the left panel of Figure \ref{fig:Evryscope_vs_TESS}. These energies are comparable, however, to the flares discovered by \citet{Schmidt_2018_ASASSN_flares} in ASAS-SN data.

Evryscope also observes the largest and rarest flare amplitudes, as displayed in the right panel of Figure \ref{fig:Evryscope_vs_TESS}. Flares emit more strongly in the blue than in the red, so our flare peak amplitude of a given flare will be several times higher than for TESS \citep{Davenport2012}.

\subsection{Most extreme superflares}\label{evr_superflares}
\begin{figure*}
	\centering
	{
		\includegraphics[trim= 1 1 1 1,clip, width=6.9in]{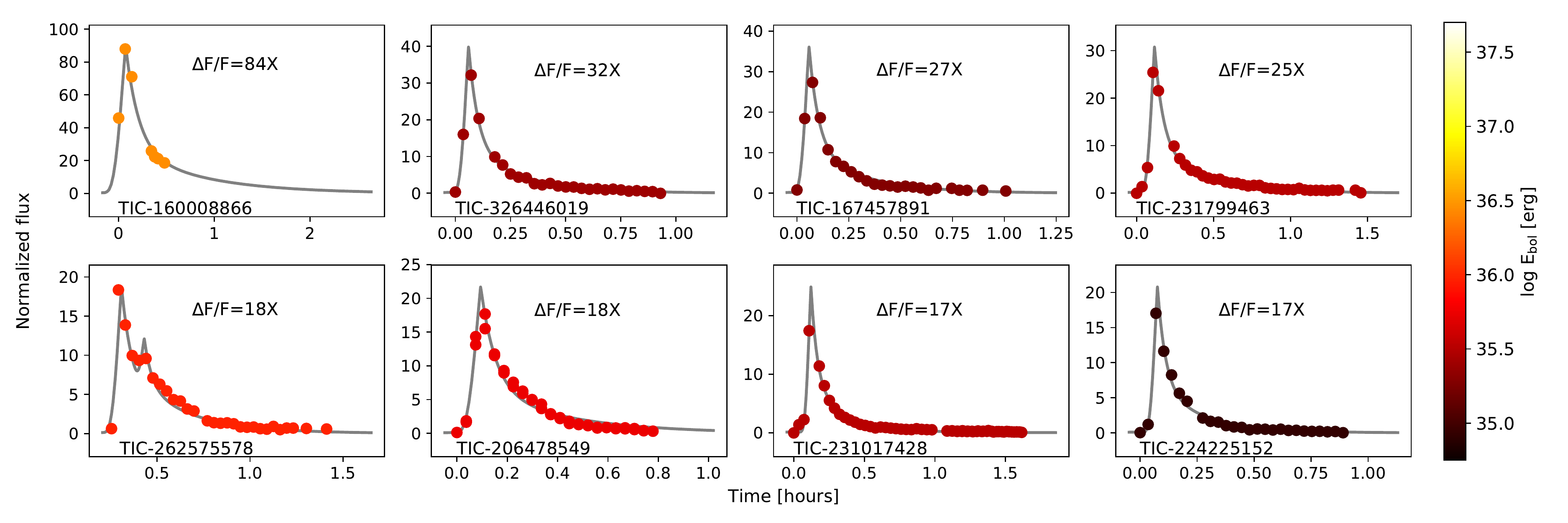}
	}
	\caption{A ``rogues gallery" of our highest-amplitude superflares detected by Evryscope from cool stars listed as 2-minute cadence-observed TESS stars in sectors 1-6. Each flare released at least 10$^{35}$ erg and is capable of significantly altering the chemical equilibrium of an Earth-like atmosphere \citep{Loyd2018}. Each flare is color-coded by its bolometric flare energy; the energy/color scheme is given on the colorbar to the right of the figure.}
	\label{fig:biggest_flares}
\end{figure*}

In 2 years of Evryscope monitoring of the nearest star, the common red dwarf Proxima Centauri, we discovered three-magnitude stellar flare events occur 2-5 times per year \citep{Howard2018}, with 2 total superflares observed \citep{Kielkopf2019}. Here, we constrain how frequently similarly-large events occur across the sky. Out of 284 flare stars, we observe 8 stellar flares that increased their star's brightness by at least 2.9 \textit{g}\textsuperscript{$\prime$}~magnitudes; they are displayed in Figure \ref{fig:biggest_flares}. These flares have also been checked against Evryscope image cutouts in addition to the regular systematics checks described in Section \ref{evr_auto_elfs}. The largest of these is a 5.6 magnitude flare from a 40 Myr M4 star in the Tuc-Hor cluster, TIC-160008866, which increased the stellar brightness by $\sim90\times$ and released 10$^{36.2}$ erg.

These superflare stars are as follows:
\begin{itemize}
  \item TIC-160008866: (UCAC2 14970156) an M4 that increased in brightness 5.6 magnitudes and released 36.2 log erg. To estimate the energy of this flare, we fit the flare template of \citet{davenport2014} and computed the area-under-the-curve. Other large flares were also observed from this star in the Evryscope light curve. Stellar activity from this young star in the Tuc Hor moving group \citep{Kraus2014} has been measured in the UV by \citet{Miles_and_Shkolnik_2017}. The extreme UV ``Hazflare" observed by \citet{HazFlare2018} is from the same cluster.
  \item TIC-326446019: (RBS 1877) an M3.5 \citep{Riaz2006} that increased in brightness 3.5 magnitudes and released 10$^{35.3}$ erg
  \item TIC-224225152: (LTT 9582) an M3 \citep{Riaz2006} that increased in brightness 3.1 magnitudes and released 10$^{34.9}$ erg
  \item TIC-231017428: (L 173-39) an M2 \citep{Gaidos2014} that increased in brightness 3.1 magnitudes and released 10$^{35.4}$ erg
  \item TIC-206478549: (WISE J035122.95-515458.1) an M4 \citep{Kraus2014} (also in the Tuc-Hor moving group) that increased in brightness 2.9 magnitudes and released 10$^{35.6}$ erg
  \item TIC-231799463: (L 57-11 B) an M4 \citep{Cowley1984} that increased in brightness 3.5 magnitudes and released 10$^{35.4}$ erg. Due to Evryscope's large pixel scale and the high PM of this system, it is possible this flare came from the M4, L 57-11 A or the semi-regular pulsator 2MASS J05125971-7027279 in the LMC \citep{Fraser2008}. All 3 stars are within $\sim$13 arcsec.
  \item TIC-262575578: (UCAC3 63-25310) an M1 that increased in brightness 3.2 magnitudes and released 10$^{35.8}$ erg
  \item TIC-167457891 (LP 767-17), an M2 that increased in brightness 3.6 magnitudes and released 10$^{35.2}$ erg.
\end{itemize}

\subsection{Superflares from TESS planet hosts}\label{TOI_flares}
Out of 284 Evryscope flare stars, one is a TESS Object of Interest (TOI). TOI-455 (TIC-98796344) was observed in TESS Sector 4, when it was found to host a candidate 1.37 R$_{\oplus}$ planet interior to the star's habitable zone (HZ). Subsequent follow-up may find a larger radius for the planet (e.g. \citet{Ziegler2018}), as another star is in the same pixel of the TESS CCD. At a distance of 20 pc, TOI-455 is close enough to make future planetary atmospheric study a possibility \citep{Ricker_TESS}. We observe a single 10$^{34.2}$ erg superflare, and predict a superflare rate of 15.1$^{23}_{-9}$ yr$^{-1}$. Although this rocky planet candidate lies outside the habitable zone, TESS is expected to discover many compact multiple-planet systems around M-dwarfs \citep{Ballard2019}. The atmospheres of any additional rocky planets in this star's HZ will also be impacted by these superflares.

\section{Astrobiological Impact of Superflares}\label{planet_habitability}
\citet{Tilley2019} find that the cumulative effect of multiple 10$^{34}$ erg superflares per year and any associated stellar energetic particles (SEPs) may destroy an Earth-like planet's ozone layer on timescales of years to decades. \citet{gunter2019} generalizes this result from \citet{Tilley2019} to estimate that a 10$^{34}$ erg superflare rate of 0.1 to 0.4 flares day$^{-1}$ is sufficient to deplete ozone. In our sample of 284 flare stars, we observe 17 flare stars in this regime.

However, the ozone loss modeling by \citet{Tilley2019} depends on the assumed distribution of particle energy versus flare energy. Efforts to directly measure the SEP environment of nearby stars by observing their stellar coronal mass ejections (CMEs) have resulted in a lack of evidence for stellar CMEs in the radio \citep{Crosley_Osten_2018a,Crosley_Osten_2018b}, although candidate CMEs have been identified in optical/X-ray data at lower SEP velocities than previously thought \citep{Moschou2019}. The lack of CMEs in the radio and reduced SEP velocities in the optical/X-ray may be due to the strong dipoles of quickly-rotating cool stars that trap SEPs before they can escape the star's magnetic field \citep{Alvarado2018}. 

We therefore inquire how many of our superflares may have sufficient energy in the UV alone to fully deplete an ozone column in a single event. \citet{Loyd2018} finds that single superflares with equivalent durations in the Si IV FUV bandpass greater than 10$^8$ seconds release enough energy to fully photo-dissociate an Earth-like planet's ozone column. \citet{Loyd2018} approximates the Si IV ED of a 3$\times$10$^{35}$ erg \textit{g}\textsuperscript{$\prime$}~-band flare to be 10$^8$ seconds. We here extend this approximation to the bolometric energy of our \textit{g}\textsuperscript{$\prime$}~flare energies. The in-band energy of an Evryscope flare is 19\% of the bolometric energy \citep{Howard2018}. As a result, a bolometric energy of 10$^{36.2}$ erg is required to exceed 10$^8$ seconds in Si IV. 

In our flare sample, we observe 1 superflare that meets this criteria. This is the 5.6 magnitude flare from the young star TIC-160008866 described in Section \ref{evr_superflares}. We also observe 23 more superflares in our sample that attain at least 10\% of our estimate of the required energy to dissociate an ozone layer. Such large flares from very young stars may not prevent planets orbiting these stars from being conducive to life. Recent modeling by \citet{OMalley2019arXiv} of the surface UV environment of Earth-analogues orbiting M-dwarfs suggests that extreme stellar activity may not prevent the formation of life, if the planet atmospheres follow the evolution of the Earth's atmosphere through time.

We note that the photo-dissociation estimates from \citet{Loyd2018} do not include modeling of the thermochemistry occurring after each flare, but rather describe how far a flare of a given energy is able to push an Earth-like atmosphere out of chemical equilibrium if the flare were to deposit its energy instantaneously. Si IV flares of 10$^8$ seconds could severely disrupt atmospheric equilibrium. During the thermochemical aftermath of such a large flare, ozone would rapidly return to equilibrium and overshoot its original value due to the creation of additional, slowly recombining free oxygen from the photolysis of O$_2$ by FUV photons. While ozone rapidly reforms after a single event, sufficiently-frequent extreme superflares would further and likely permanently disrupt atmospheric equilibrium.

We also note that extreme UV radiation and high energy SEPs from superflares will alter planetary atmospheric chemistry and surface environments through more pathways than ozone depletion. For example, the atmospheric volatile composition of close-in planets may be altered by SEPs associated with superflares through the production of secondary particles. These SEPs would also increase the surface radiation dosage, although potentially not to un-inhabitable levels \citep{Atri2017}. As a second example, SEPs from superflares may fix inert atmospheric nitrogen in Earth-like atmospheres, creating greenhouse gasses and compounds necessary for life \citep{Airapetian2016}.

Although an Earth-like atmosphere may not survive repeated flaring, many HZ planets may orbit inactive stars. During the 2-year primary TESS mission, planets as small as 2R$_{\oplus}$ and 1.6R$_{\oplus}$ may be detected within the HZ of only 1822 and 1690 stars, respectively \citep{Kaltenegger2019}. We observe a total of 49 stars in the TESS HZ catalog to exhibit large flares. Due to the faintness of many of these catalog stars in the blue, we only search 335 catalog stars, for a physical rate of 14.6$\pm$2\% with large flares. 

\section{Conclusions}\label{conclude}
Approximately two-thirds of cool stars are active \citep{west2015}, raising concerns about the habitability of the planets orbiting many of these stars, which make up the majority of the Galaxy's stellar population. As TESS searches for Earth-sized planets around these active host stars, constraining their superflare occurrence remains a key step in assessing potential habitability. Evryscope has performed long-term high-cadence monitoring of every bright Southern TESS planet-search target. With this data, we record the long-term superflare rates of 4068 cool flaring stars observed by TESS in its first six sectors.

We observe 575 flares from 284 flare stars, with a marked increase in flaring at spectral types close to the M4 fully-convective boundary. We find a decrease in average flare energy at later spectral types arising from the decreasing size of the stellar convective region. We present average FFDs of active stars as a function of spectral type and measure the annual superflare rates of each spectral type, with late-K and early-M dwarfs demonstrating the highest rates. We also find that the largest flare amplitudes expected from a flaring star of each spectral type in a given observation time increases for later types. We find that the decay times of our superflares are broadly consistent with emission caused by magnetic re-connection, although we cannot rule out the possibility of further emission mechanisms. We approximate the minimum starspot coverage required to produce superflares, and hypothesize the maximum allowed values of superflare energy and waiting time between flares corresponding to 100\% hemispherical spot coverage. Such values are extrapolations from the G-dwarf superflare regime and should be treated with caution, especially since the minimum spot area we compute may be at least an order-of-magnitude less than the true spot area. Finally, we observe a decreasing superflare rate for older stars at high galactic latitude.

Among our superflare sample, we observe a number of extreme events. We observe 8 flares that increased the brightness of their host star by at least 3 stellar magnitudes in \textit{g}\textsuperscript{$\prime$}~and released at least 10$^{35}$ erg. The largest of these flares is a 5.6 magnitude event from an active 40 Myr-old Tuc-Hor cluster member. This flare released 10$^{36.2}$ erg, enough energy to completely photo-dissociate the ozone column of an Earth-like planet in one event. If we factor in high energy particles potentially associated with flares, lesser superflares become equally dangerous. For example, we find 17 stars that may fully attenuate an Earth-like atmosphere via repeated flaring by emitting at least 0.1 10$^{34}$ erg flares d$^{-1}$. Of the 1822 stars around which TESS may discover planets smaller than $2\mathrm{R}_{\oplus}$ in the HZ, we observe only 49 to emit large flares. Because most of these 1822 host stars are faint in the blue, we only searched the brightest 335 for flares, resulting in 14.6$\pm$2\% with large flares.

We also observe a 10$^{34}$ erg superflare from the mid M-dwarf TOI-455 (TIC-98796344). Host to a nearby 1.4R$_{\oplus}$ planet candidate interior to the habitable zone, the atmosphere of a planet orbiting TOI-455 may be suitable for future study. We constrain the superflare rate of this TOI to be 15.1$^{23}_{-9}$ yr$^{-1}$. Even if the radius is found to be larger (but still non-stellar) as a result of dilution from nearby stars, its atmosphere may be altered by superflares and associated SEPs. Future work obtaining transit spectroscopy of TOI-455 or other flaring host stars to a transiting planet within months of an Evryscope-detected superflare may enable constraints on changes to a planetary atmosphere.

Upon future publication of flares across the entire Southern sky, the Evryscope sample of superflares will more than double (i.e. adding the subset of flares from 7 new TESS sectors) and the number of flares discoverable in TESS light curves will likely increase by a factor of 6$\times$ (i.e. adding flares from 11 new sectors) ( \citet{gunter2019}, \citet{gunter2019b}, in preparation). More work is needed to analyze TESS and Evryscope flares from each star observed by both surveys. By combining the frequently-occurring small and moderate flares seen by TESS across 28 days with rare superflares observed over multiple years by Evryscope, we may better explore the FFD of each star in the South. From a well-constrained FFD, planetary atmosphere modeling for rocky TESS planets orbiting flare stars will inform the atmospheric compositions and surface UV environments of these worlds. Well-constrained FFDs for such a large sample will also make possible large-scale statistical treatments of superflare occurrence as functions of stellar rotation, stellar age, binarity, and surface magnetic field topology. Multi-band Evryscope plus TESS superflares observed at high cadence in both the red and blue will also inform the temporal evolution of the flare blackbody and plasma environment for these events (e.g \citet{Kowalski2016_ULTRACAM}).

Because our sample of flare stars have both Evryscope and TESS light curves, rotation period measurements from both surveys may be combined and then compared against the flare parameters of each star to search for changes in cool star spin-down and magnetic-field evolution (e.g. \citet{Mondrik2019}). We urge further work in this area.

\section*{Acknowledgements}\label{acknowledge}
We would like to thank the anonymous referee who graciously gave their time to make this the best version of this work.

The authors wish to thank Andrew Mann, Parke Loyd, James Davenport, and Evgenya L. Shkolnik for insightful comments.  WH, HC, NL, JR, and AG acknowledge funding support by the National Science Foundation CAREER grant 1555175, and the Research Corporation Scialog grants 23782 and 23822. HC is supported by the National Science Foundation Graduate Research Fellowship under Grant No. DGE-1144081. OF and DdS acknowledge support by the Spanish Ministerio de Econom\'{\i}a y Competitividad (MINECO/FEDER, UE) under grants AYA2013-47447-C3-1-P, AYA2016-76012-C3-1-P, MDM-2014-0369 of ICCUB (Unidad de Excelencia `Mar\'{\i}a de Maeztu'). The Evryscope was constructed under National Science Foundation/ATI grant AST-1407589.
\par This paper includes data collected by the TESS mission. Funding for the TESS mission is provided by the NASA Explorer Program.
\par This research has made use of the Exoplanet Follow-up Observation Program website, which is operated by the California Institute of Technology, under contract with the National Aeronautics and Space Administration under the Exoplanet Exploration Program.
\par This work has made use of data from the European Space Agency (ESA) mission {\it Gaia} (\url{https://www.cosmos.esa.int/gaia}), processed by the {\it Gaia} Data Processing and Analysis Consortium (DPAC, \url{https://www.cosmos.esa.int/web/gaia/dpac/consortium}). Funding for the DPAC has been provided by national institutions, in particular the institutions participating in the {\it Gaia} Multilateral Agreement.
This research made use of Astropy,\footnote[2]{http://www.astropy.org} a community-developed core Python package for Astronomy \citep{astropy:2013, astropy:2018}, and the NumPy, SciPy, and Matplotlib Python modules \citep{numpyscipy,Jones2001,matplotlib}.

{\it Facilities:} \facility{CTIO:Evryscope}, \facility{TESS}

\bibliographystyle{apj}
\bibliography{paper_references}


\end{document}